\documentclass[%
reprint,
 amsmath,amssymb,
 aps,
]{revtex4-2}
\usepackage{graphicx}
\usepackage{dcolumn}
\usepackage{bm}
\usepackage{hyperref}
\usepackage{appendix}
\usepackage{amsmath}
\usepackage{multirow}
\usepackage{tabularx}
\usepackage{booktabs}
\usepackage{caption}
\usepackage{float}
\usepackage{IEEEtrantools}
\usepackage{subcaption}

\usepackage[labelformat=empty]{subcaption}
\usepackage[section]{placeins} 
\usepackage{caption} 
\hypersetup{hypertex=true,
            colorlinks=true,
            linkcolor=blue,
            anchorcolor=blue,
            citecolor=blue,
            urlcolor=blue}

\begin{document}

\preprint{APS/123-QED}

\title{Microscopic Study of Charge Properties in Halo Nuclei}
\author{Yun Dong Wang}
 \affiliation{College of Physics, Jilin University, Changchun 130012, China}
\author{Hui Hui Xie}
 \affiliation{College of Physics, Jilin University, Changchun 130012, China}
\author{Tian Shuai Shang}
 \affiliation{College of Physics, Jilin University, Changchun 130012, China}
\author{Peng Xiang Du}
 \affiliation{College of Physics, Jilin University, Changchun 130012, China}
\author{Jian Li}\email{jianli@jlu.edu.cn}
\affiliation{College of Physics, Jilin University, Changchun 130012, China}
\email{E-mail:jianli@jlu.edu.cn}
\author{Haozhao Liang}
\affiliation{Department of Physics, Graduate School of Science,	The University of Tokyo, Tokyo 113-0033, Japan}
\affiliation{Quark Nuclear Science Institute, The University of Tokyo, Tokyo 113-0033, Japan}
\affiliation{RIKEN Center for Interdisciplinary Theoretical and Mathematical Sciences (iTHEMS), Wako 351-0198, Japan}
\author{Kaiyuan Zhang}
\affiliation{National Key Laboratory of Neutron Science and Technology, Institute of Nuclear Physics and Chemistry, China Academy of Engineering Physics, Mianyang, Sichuan, 621900, China}
%



\date{\today}

\begin{abstract}

Employing the relativistic continuum Hartree-Bogoliubov (RCHB) theory with intrinsic electromagnetic structure corrections, this work primarily investigates the charge properties of halo nuclei along the Ne and P isotopic chains. Our results characterize halo nuclei by an extended tail in the charge density and distinct signatures in the charge form factors at low momentum transfer. Moreover, the higher-order radial moments of nuclear charge density, particularly the eighth moment, exhibit pronounced oscillations, serving as a key indicator of the neutron halo structure. These special features in nuclear charge distributions will serve as key references for identifying halo nuclei.

\end{abstract}

\maketitle


\section{\label{sec:1}introduction}

The study of halo nuclei is of fundamental importance in modern nuclear physics, as it directly challenges and extends conventional paradigms of nuclear structure such as the $A^{1/3}$ radius law, which is the foundation of the collective model \cite{RMP76215}. This exotic phenomenon, often arising from the breakdown of the conventional shell model \cite{plb707357}, manifests as a diffuse neutron or proton halo surrounding a tightly bound core \cite{prl80460,prc041302,prc064313}. These systems are characterized by their weak binding and continuum coupling \cite{RMP76215,prl80460,PPNP68215}, offering a unique quantum laboratory for exploring few-body correlations in open systems. Investigating these anomalous nuclei is thus crucial for developing a more complete understanding of nuclear forces and structures far from stability.\par

The study of halo nuclei is primarily motivated by their exceptionally large matter radius compared to stable nuclei \cite{prl552676}. Experimentally, this extended structure can be inferred from several reaction observables. For instance, enhanced interaction cross-sections, narrow momentum distributions in fragmentation or knockout reactions \cite{plb707357,PPNP68215,prl242501,prl222504,prl262501,prl142501}, and a significant concentration of electric dipole ($E1$) strength in the low-excitation energy region, often referred to as soft dipole excitation \cite{prl831112,plb39411,prl252502}, all indicate a diffuse matter density distribution. The emergence of halo structures is closely linked to the low separation energy of the valence nucleon(s) and their occupation of low-angular-momentum orbitals, which serves as a necessary structural condition for halo formation \cite{PPNP68215,prl142501,prl242501}. Recently, the study of mirror-symmetry breaking, probed by comparing halo nuclei with their mirror partners to extract Coulomb energy differences arising from asymmetric proton distributions, provides another possible probe of proton halo structure \cite{prl222501}.\par

The theoretical investigation of halo nuclei builds on several well-established approaches, including density functional theory (DFT) \cite{npa600371,prc064305,prl773963,prl80460}, effective field theory \cite{prc0143252,prc044004,JPG103002,prc044325,prc024318,prc044304,prc014001}, and $ab\;initio$ methods \cite{prc034305,prl2425012,prl2225012,prc061302,prcL061304,prc014306,prl162503}. Among these, the relativistic density functional theory (RDFT) provides a particularly powerful framework \cite{PPNP57470}. Its implementations, such as the relativistic Hartree–Bogoliubov (RHB) approach—which incorporates the Bogoliubov transformation into the relativistic mean-field framework, have been successfully applied to the study of both spherical and deformed halo systems \cite{prl773963,prl80460,prc011301,plb138112,prcL041301,plb138792,prc044308,plb139989,plb785530,plb138422,epja60251}. Such theoretical studies have clarified the relationship between halo formation and shell evolution, and provided a self-consistent treatment of the continuum in halo nuclei. The relativistic continuum Hartree-Bogoliubov (RCHB) theory, in particular, has achieved significant success in describing halo nuclei. It effectively identifies halo configurations by analyzing the single-particle levels and their occupation probabilities in the canonical basis, which can reveal whether valence nucleons occupy weakly bound, low-angular-momentum orbits \cite{prl773963,prl80460,cpc3643,prc041302}. Notably, it successfully described the halo phenomenon in $^{11}\mathrm{Li}$ in 1996 \cite{prl773963} and later predicted the giant halo phenomenon \cite{prc041302,prl80460}.\par

While the investigation of halo nuclei mainly focuses on elucidating the underlying mechanisms of their formation, their charge properties represent another critical aspect for a complete structural characterization. Existing studies have extensively explored properties such as charge density distributions, charge radii, and charge form factors in halo nuclei \cite{plb775126,prc057301,prc024318,prc014325,prc044307,plb139082}. Notably, the study of 2$p$ halo nuclei ($A = 18–34$) reveals a diffuse tail in the charge density distribution, leading to a significantly larger charge radius in proton halo nuclei \cite{plb775126}. In neutron halo nuclei, a more subtle increase in charge radius may also occur due to center-of-mass motion and polarization of the core \cite{prc057301}. These increases in charge radius and diffuse tail in the charge density distribution directly manifest in the charge form factor, which exhibits characteristical difference at low momentum transfer compared to stable nuclei \cite{prc024318,prc014325,prc044307}.
These charge properties play a crucial role in understanding halo phenomena, yet a quantitative understanding of them remains elusive. Previous studies have often focused on specific observables, providing valuable insights but fall short of a unified quantitative picture. Furthermore, many calculations still rely on the conventional point-proton approximation \cite{plb775126,prc044307,adndt101488,adndt101661}, where the point-proton density is directly equated with the nuclear charge density distribution, thereby neglecting intrinsic electromagnetic structure corrections such as the finite size of nucleons and the contributions of nucleon spin-orbit densities to the charge density. Studies have shown that these corrections significantly affect the calculations of both charge and weak-charge radii \cite{prc045503} and may be particularly important in light halo nuclei \cite{prc014320}. Therefore, to achieve a consistent and precise quantitative description, it is essential to move beyond the point-proton approximation and adopt a more refined treatment of the charge density.\par

To this end, the present study employs the RCHB theory, further developed to incorporate intrinsic electromagnetic structure corrections. This improved framework has demonstrated its reliability by accurately reproducing the charge radii of Ca and Pb isotopes \cite{prc064319,prcL021303}. Building on this foundation, we investigate the charge properties of halo nuclei along the Ne and P isotope chains, including their charge density distributions, charge radii, charge form factors, and higher-order charge moments. Additionally, the evolution of the eighth-order charge moment with neutron number will be analyzed for isotopes of Ne, Mg, F, and Ca. Although the RCHB framework assumes spherical symmetry, it could, to some extent, account for key features of the halo structure in deformed nuclei \cite{cpc3643}. Given that most halo nuclei are deformed, the inclusion of deformation effects is essential for a more sufficient and consistent description. Therefore, future work will extend this investigation by employing the deformed relativistic Hartree-Bogoliubov theory in continuum (DRHBc) \cite{prc011301,prc024312,prc024314,prc014316} with the intrinsic electromagnetic structure corrections. Furthermore, recent progress has been made in the application of deep neural networks for precisely predicting nuclear charge densities \cite{nst33153,prc014308,nst3793}, which is expected to provide valuable insights for subsequent investigations into the charge properties of halo nuclei.\par

The fundamental formalism of the RCHB method and the procedure for constructing nuclear charge properties through RCHB theoretical calculations with self-consistent intrinsic electromagnetic structure corrections are presented in Sec. \ref{sec:2}. The corresponding results and related discussions are provided in Sec. \ref{sec:3}, followed by a summary of the work and an outlook for future research in Sec. \ref{sec:4}.\par


\section{\label{sec:2}THEORETICAL FRAMEWORK}

\begin{figure*}
\centering
\includegraphics[width=16cm]{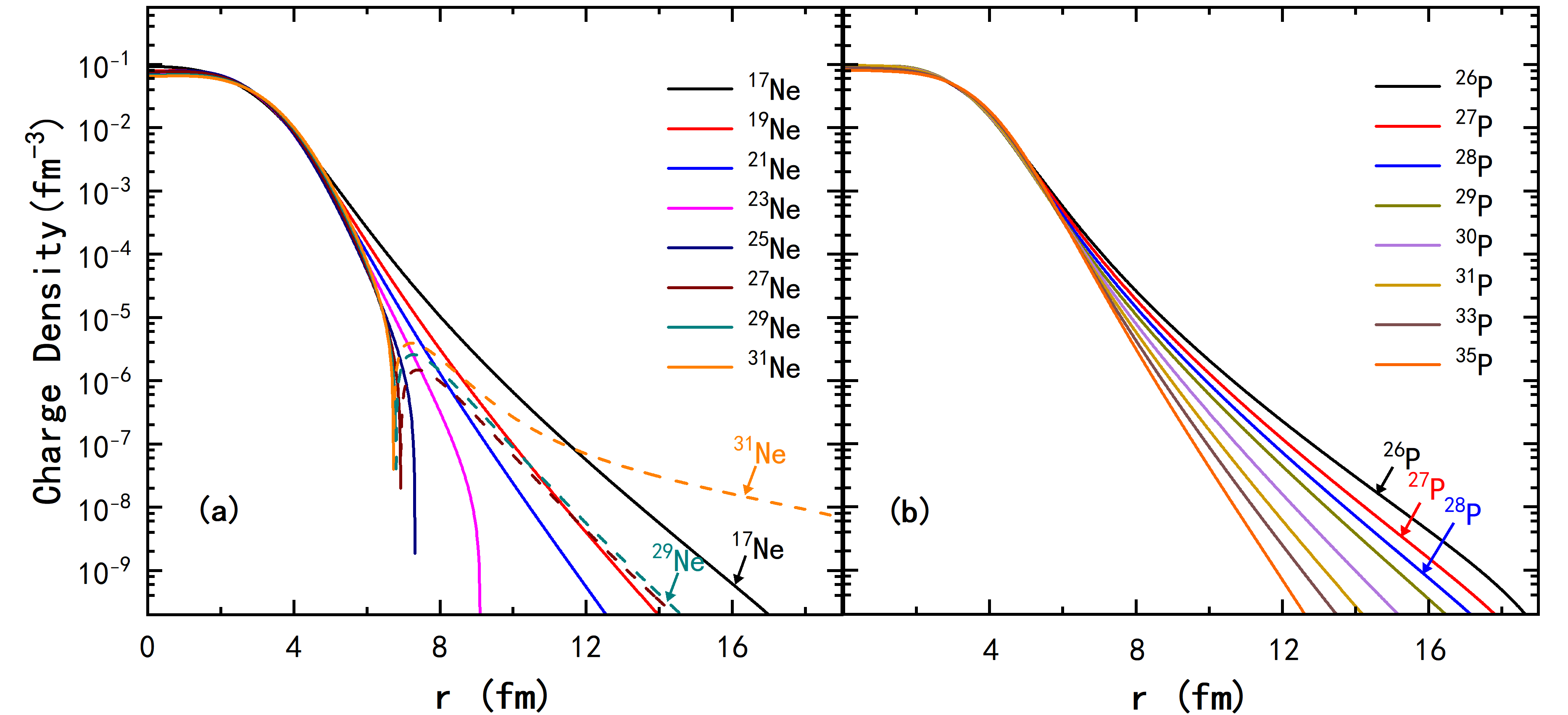}
\caption{\raggedright\label{fig:wide}Charge density distributions for the Ne and P isotopic chains, obtained from the RCHB theory with intrinsic electromagnetic structure corrections. The dashed lines in (a) indicate the absolute values of the negative charge densities. The nuclei $^{31}\text{Ne}$ (known neutron halo) and $^{17}\text{Ne}$ (known proton halo) are shown,  along with the neutron halo candidate $^{29}\text{Ne}$ and the proton halo candidates $^{26,27,28}\text{P}$ \cite{plb707357,prl262501,npa635292,plb57121,plb707357,prl815089}.}\label{fig:1}
   \subcaptionbox{\label{fig:1(a)}}{}
   \subcaptionbox{\label{fig:1(b)}}{}
\end{figure*}

\subsection{Relativistic continuum Hartree-Bogoliubov method}
 The theoretical construction begins with the Lagrangian density, from which the energy density functional of the nuclear system is established under the mean-field and no-sea approximations. Minimization of this energy density functional with respect to the densities yields the RHB equation \cite{zpa33923}, 

\begin{equation}
    \begin{pmatrix}
h_D - \lambda & \Delta \\
-\Delta^* & -h_D^* + \lambda
\end{pmatrix}
\begin{pmatrix}
U_k \\ V_k
\end{pmatrix}
= E_k
\begin{pmatrix}
U_k \\ V_k
\end{pmatrix},\label{eq:0}
\end{equation}
where $E_k$ is the quasiparticle energy, $U_k$ and $V_k$ are the upper and lower components of the quasiparticle wave function, respectively, $\lambda$ are the
chemical potentials for nucleons, $h_D$ refers to
the Dirac Hamiltonian,
\begin{equation}
    h_D(\boldsymbol{r}) = \boldsymbol{\alpha}\cdot\boldsymbol{p}+V(\boldsymbol{r})+\beta[M+S(\boldsymbol{r})],\label{eq:hd}
\end{equation}
where $S(\boldsymbol{r})$ and $V(\boldsymbol{r})$ are the scalar and vector potentials, respectively. $\Delta$ is the pairing potential, with a density-dependent force of zero range.\par

For weakly bound nuclei with Fermi surfaces close to the threshold, the possible continuum coupling induced by pairing correlations might lead to diffusion of nuclear densities \cite{prc024314}. Solving the RHB equations in coordinate space \cite{npa6353} or in a Woods-Saxon basis \cite{prc034323,prc024302} could well account for such continuum effects via the preservation of a proper asymptotic behavior of wave functions. In this framework, the point-proton and point-neutron densities are defined within the RCHB theory as the probability distributions of point-like nucleons. They are given by the quasiparticle wave functions obtained from the self-consistent solution of the RHB equations:
\begin{equation}
    \rho_\tau(r) =\sum_{k \in \tau}\frac{n_k}{4\pi r^2}\{[G_V^k(r)]^2+[F_V^k(r)]^2\},\label{eq:1}
\end{equation}
where $n_k$ represents the occupation probability of the $k$-th orbital, while $G_V^k(r)$ and $F_V^k(r)$ correspond to the large and small components of the quasiparticle wave functions, respectively. In this work, three point-coupling density functionals, PC-F1 \cite{prc0443082}, PC-PK1 \cite{prc054319}, and PC-L3R \cite{plb137946}, were employed. Similar numerical results and fully consistent conclusions were obtained with these three functionals. Among them, PC-PK1 has been proven successful in describing nuclear ground-state properties and is recognized as one of the best density functionals for nuclear properties \cite{prcL021301,prc014301,AAPPSbull3513,nst36231,adndt101488}. The key numerical parameters adopted in our calculations, including the box size $R_\text{box}=20 \;\text{fm}$, the radial step size $x_\text{step}=0.1 \;\text{fm}$, and the angular momentum cutoff $J_\text{max}=19/2 \;\hbar$, are exactly the same as those used in Ref. \cite{adndt1211}. In that reference, the authors performed detailed checks and discussions on the convergence of these numerical parameters within the RCHB theory with the PC-PK1 functional. Given that the present work focuses on halo nuclei with very extended density distributions, we have performed additional convergence checks for the charge density, charge radius, and higher-order moments of the charge density. These checks are conducted by varying the box size from 16 fm to 24 fm while adopting a small pairing strength of $–150\;\text{MeV}\cdot \text{fm}^3$. This avoids renormalizing the zero-range pairing strength with respect to the corresponding model space and maintains the convergence of self-consistent iterations. The tail of the charge density distribution obtained with $R_\text{box} = 20\;\text{fm}$ and that with $R_\text{box} = 24\;\text{fm}$ remain within the same order of magnitude. The relative difference in the charge radius is less than $0.0032\%$, and that in the fourth-order moment is below $0.23\%$. The eighth-order moment also does not change the qualitative conclusions of this work, which will be discussed in detail below. Further details of these checks are presented in the Appendix. With these validations established, the following discussion will focus on the results obtained with the PC-PK1 functional under the same numerical conditions as those in Ref. \cite{adndt1211}, where the pairing strength of $–342.5\;\text{MeV}\cdot \text{fm}^3$ is determined by experimental odd-even mass differences. Further theoretical details can be found in Ref. \cite{npa6353}.

\subsection{Nuclear charge properties}

The nuclear charge form factor $ F_c$, which is the Fourier transform of the charge density in momentum space, is derived from the single-particle electromagnetic current operator \cite{PPNP59694} as the ground-state expectation value of its zero component. It encompasses the convolutions of the point-nucleon spatial distributions with their intrinsic Sachs electric form factors, as well as the contributions from the nucleon spin-orbit densities described by the Pauli form factors. The complete expression reads:

\begin{equation}
    F_c(\bm{q}) = \int d^3 \bm{r} e^{i\bm{q}\cdot \bm{r}}\sum_{\tau \in \{p,n\}}[G_{E\tau}(q^2)\rho_\tau(r)+F_{2\tau}(q^2)W_\tau(\bm{r})],\label{eq:2}
\end{equation}
where $G_{E\tau}$ and $F_{2\tau}$ denote the Sachs electric and Pauli nucleon form factors, respectively. The point-nucleon density $\rho_{\tau}$ is derived from Eq. (\ref{eq:1}), and the general form of the spin-orbit density $W_\tau(r)$ is given by \cite{prc054303,PTEP86127}

\begin{equation}
    W_\tau(\bm{r}) = \frac{\mu_\tau}{2M}(-\frac{\nabla^2\rho_\tau(r)}{2M}+i\nabla\cdot \langle 0|\sum_{k \in\tau}\delta(\bm{r}-\bm{r}_k)\bm{\gamma}_k|0 \rangle),\label{eq:3}
\end{equation}
where $\mu_\tau$ and $\gamma_k$ are the anomalous magnetic moment and the $\gamma$ matrix of the $k$th nucleon, respectively.\par

The nuclear charge density is related to $F_c$, which represents the charge density distribution in momentum space. Performing an inverse Fourier transformation on $F_c$ yields the nuclear charge density in coordinate space, which incorporates the intrinsic electromagnetic structure corrections and takes the form

\begin{equation}
    \rho_c(r) =\sum_{\tau \in \{p,n\}}[\rho_{c\tau}(r)+W_{c\tau}(r)],\label{eq:4}
\end{equation}
where $\rho_{c\tau}(r)$ and $W_{c\tau}(r)$ represent the contribution of the nucleon density and spin-orbit density to charge density, respectively. To account for the finite size of the nucleons, the former is obtained by folding the point-nucleon density $\rho_\tau$ with the nucleonic charge density \cite{prc054303}, more details are demonstrated in Ref. \cite{prA042807}.\par

With the corrected charge density (including intrinsic electromagnetic structure effects) determined, the higher-order moments of the nuclear charge density can be calculated directly,

\begin{equation}
R^{n}_\text{ch}\equiv \langle r^n\rangle_c=\frac{4\pi}{Z}\int_0^\infty \rho_c(r) r^{n+2}dr.\label{eq:5}
\end{equation}
In particular, the second-order moment of charge density \cite{PTEP86127} is
\begin{equation}
\begin{aligned}
R_\text{ch}^2 \equiv \langle r^2 \rangle_c &=\frac{4\pi}{Z}\int_0^\infty \rho_c(r) r^{4}dr \\
       &=\langle r^2 \rangle_p+r_p^2+\langle r^2 \rangle_{W_p}+\frac{N}{Z}(r^2_n+\langle r^2 \rangle_{W_n}),
\end{aligned}
\label{eq:6}
\end{equation}
where $R_\text{ch}$ is the charge radius, $\langle r^2 \rangle_p$ is the second moment of the point-proton density; $\langle r^2 \rangle_{W_p}$ and $\langle r^2 \rangle_{W_n}$ are the second moments of the spin-orbit densities for the proton and neutron, respectively; $r_p=0.8414\;\text{fm}$ \cite{RMP025010} and $r_n^2=-0.11\;\text{fm}^2$ \cite{nc121759} denote the charge radius of the proton and the mean-square radius of the neutron, which account for the ﬁnite-size effect of the nucleon.

\section{\label{sec:3}Results and discussion}

\begin{figure}
\centering
\includegraphics[width=9cm]{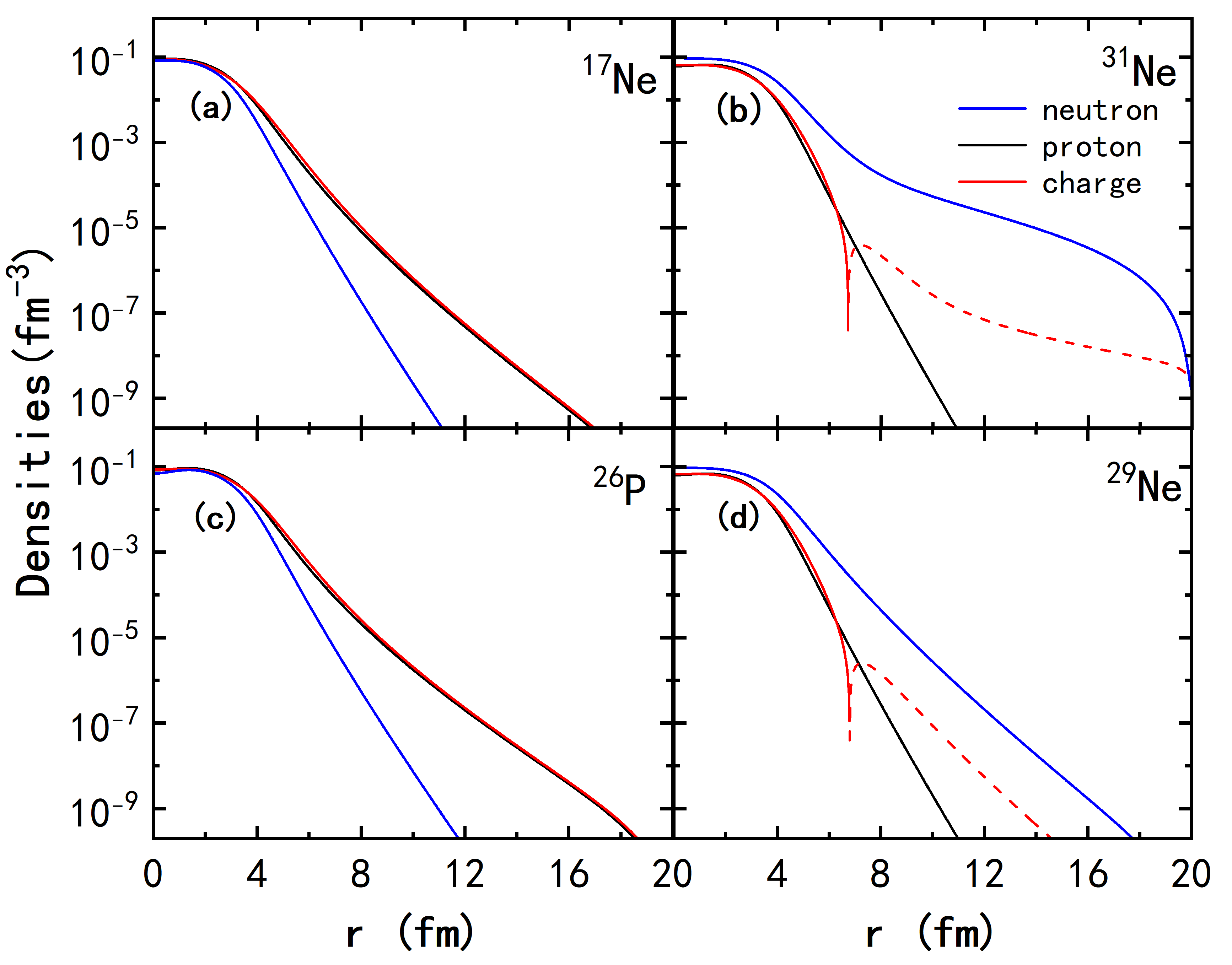}
\caption{\raggedright\label{fig:wide}Point-neutron (blue), point-proton (black), and charge (red) density distributions for $^{17}\text{Ne}$, $^{31}\text{Ne}$, $^{26}\text{P}$, and $^{29}\text{Ne}$, obtained from the RCHB theory with intrinsic electromagnetic structure correction. In panels (b) and (d), the dashed red line shows the absolute value of the negative charge density.}\label{fig:2}
   \subcaptionbox{\label{fig:2(a)}}{}
   \subcaptionbox{\label{fig:2(b)}}{}
   \subcaptionbox{\label{fig:2(c)}}{}
   \subcaptionbox{\label{fig:2(d)}}{}
\end{figure}

\begin{table*}[t] 
    \caption{\label{tab:table1}%
Point-proton and point-neutron radii and charge radii, and individual contributions of the electromagnetic structure corrections, obtained from RCHB calculations.
    }
    \begin{ruledtabular}
        \begin{tabular}{ccccccc}
             & \(R_\text{p}\)(fm) & \(R_\text{n}\)(fm) & \(R_\text{ch}\)(fm) & \(\langle r^2 \rangle_\text{wp} \)($\text{fm}^2$) & \( \frac{N}{Z}\langle r^2 \rangle_\text{wn} \)($\text{fm}^2$) & \(\frac{N}{Z} r^2 _\text{n} \)($\text{fm}^2$)\\
             \hline
             $^{17}\text{Ne}$ & 2.89114 & 2.55299 & 2.99905 & 0.04336 & $-$0.03868 & $-$0.07700  \\
             $^{29}\text{Ne}$ & 2.89186 & 3.40191 & 2.97557 & 0.05857 & $-$0.06637 & $-$0.20900  \\
             $^{31}\text{Ne}$ & 2.92585 & 3.82749 & 3.00832 & 0.05798 & $-$0.04578 & $-$0.23100  \\
             $^{26}\text{P}$  & 3.12375 & 2.81731 & 3.22841 & 0.09786 & $-$0.06037 & $-$0.08067  \\
             $^{27}\text{P}$  & 3.08608 & 2.85335 & 3.18895 & 0.10060 & $-$0.07506 & $-$0.08800  \\
             $^{28}\text{P}$  & 3.03922 & 2.87190 & 3.13952 & 0.10441 & $-$0.09732 & $-$0.09533  \\ 
             
        \end{tabular}
    \end{ruledtabular}
\end{table*}

\begin{figure*}
\centering
\includegraphics[width=16cm]{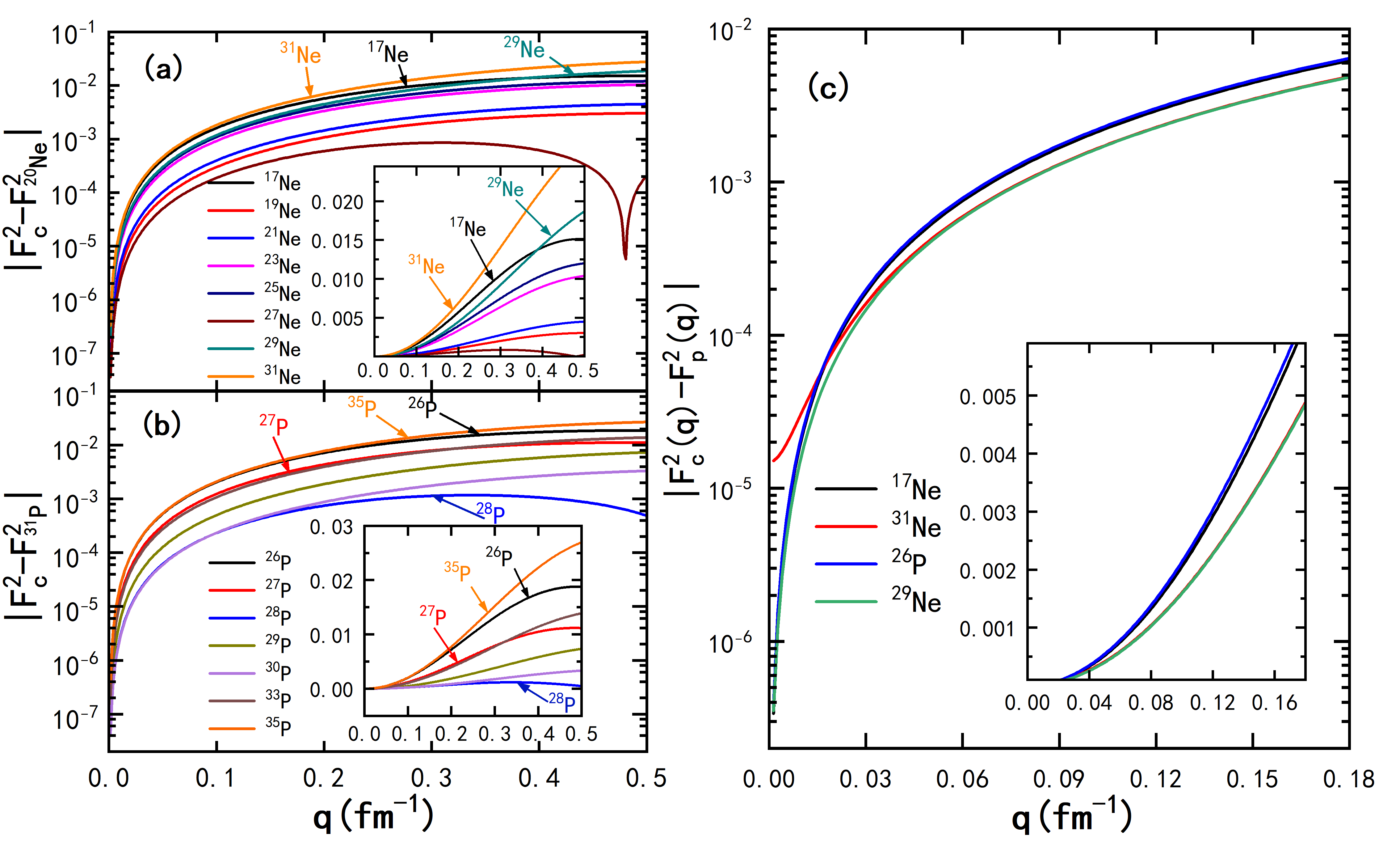}
\caption{\raggedright\label{fig:wide}Absolute differences in the squared charge form factor at low momentum transfer. (a) shows results for P isotopes relative to the reference nucleus \(^{31}\mathrm{P}\); (b) for Ne isotopes relative to \(^{20}\mathrm{Ne}\); and (c) for \(^{17}\mathrm{Ne}\), \(^{31}\mathrm{Ne}\), \(^{26}\mathrm{P}\), and \(^{29}\mathrm{P}\) relative to the form factors of their respective point-proton density distributions.}\label{fig:3}
   \subcaptionbox{\label{fig:3(a)}}{}
   \subcaptionbox{\label{fig:3(b)}}{}
   \subcaptionbox{\label{fig:3(c)}}{}
\end{figure*}

\begin{figure}
\centering
\includegraphics[width=9cm]{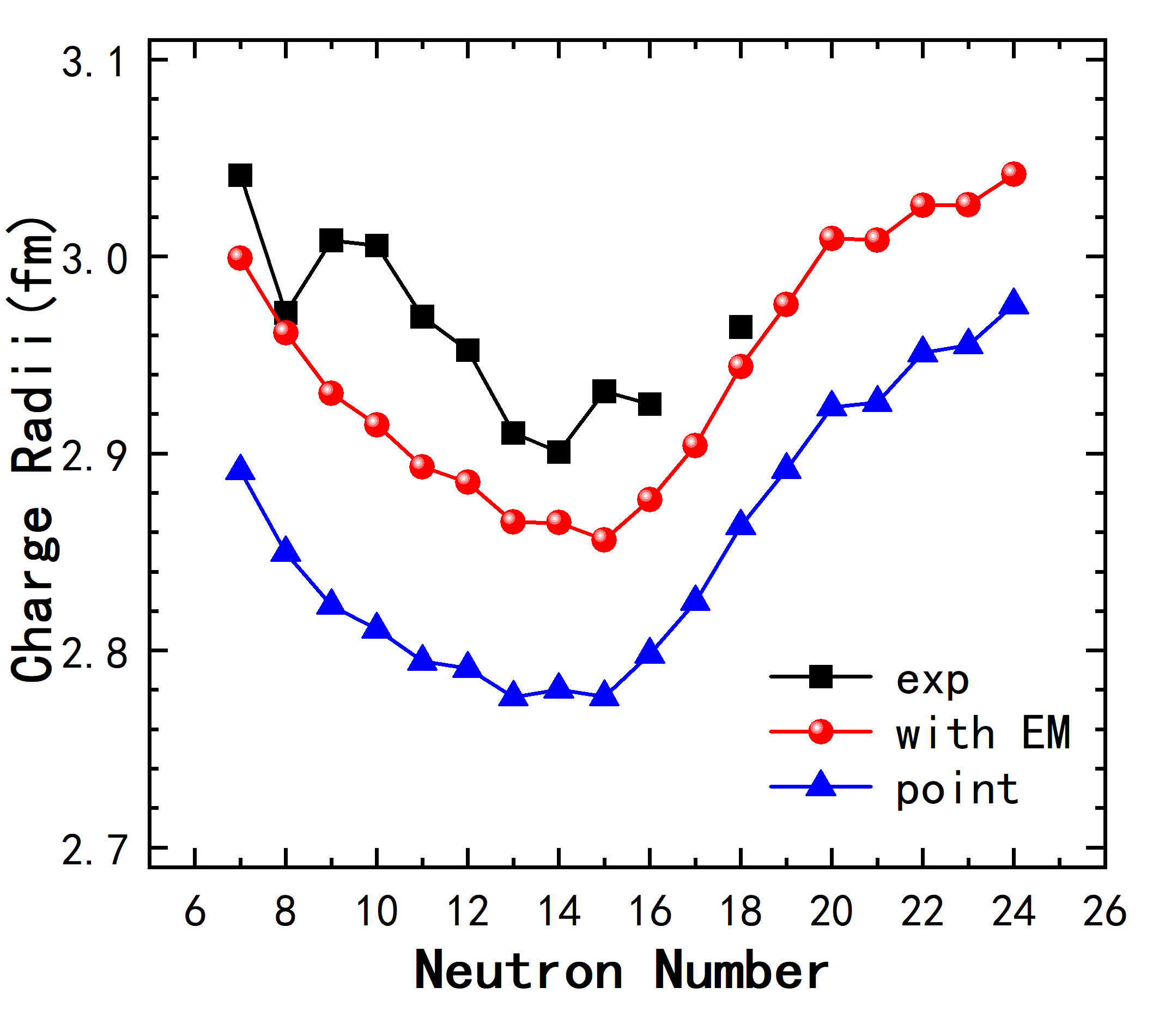}
\caption{\raggedright\label{fig:wide}Comparison of experimental charge radii for Ne isotopes with the calculated point-proton radii and the charge radii incorporating intrinsic electromagnetic structure corrections (with EM), obtained from RCHB calculations.}\label{fig:o}
\end{figure}

\begin{figure}
\centering
\includegraphics[width=9cm]{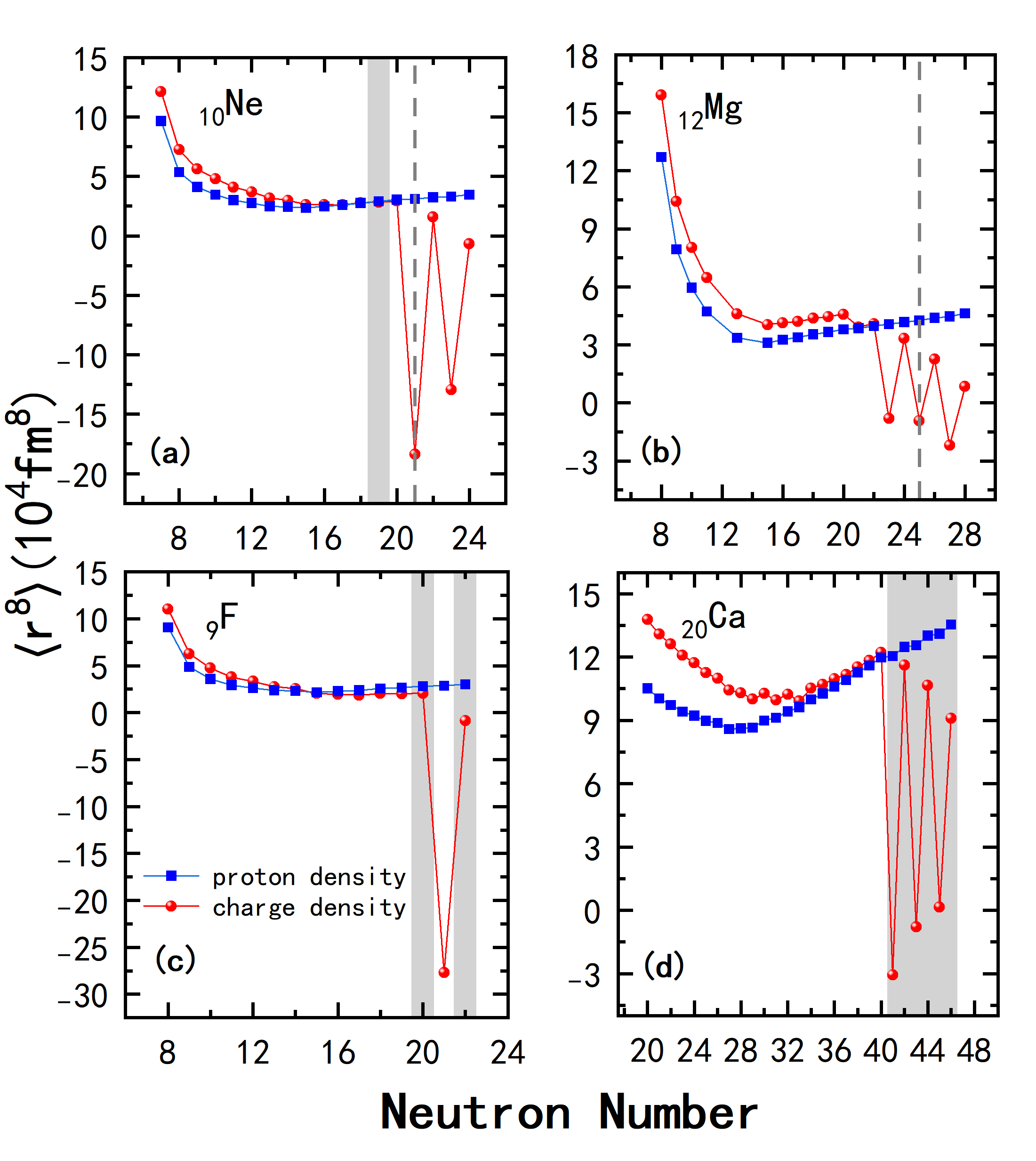}
\caption{\raggedright\label{fig:wide}Comparison of the eighth-order charge radial moments for the F, Ne, Mg, and Ca isotopic chains. The blue point denotes the point-proton density, while the red represents the charge density. The dashed segments indicate confirmed halo nuclei, and the gray shaded areas mark predicted halo nuclei. \cite{plb707357,prl262501,npa635292,prl242501,prl222504,prc014328,prc041302}. }\label{fig:4}
   \subcaptionbox{\label{fig:4(a)}}{}
   \subcaptionbox{\label{fig:4(b)}}{}
   \subcaptionbox{\label{fig:4(c)}}{}
   \subcaptionbox{\label{fig:4(d)}}{}
\end{figure}

\begin{table*}[t] 
    \caption{\label{tab:table2}%
The second, fourth, sixth, and eighth-order moments of the charge density for Ne and P isotopic chains.
    }
    \begin{ruledtabular}
        \begin{tabular}{cccccccccccc}
             &\(R^2_\text{ch(exp)}(\text{fm}^2)\) \cite{adndt9969}& \(R^2_\text{ch}(\text{fm}^2)\) & \(R^4_\text{ch}(\text{fm}^4)\) & \(R^6_\text{ch}(\text{fm}^6)\) & \(R^8_\text{ch}(\text{fm}^8)\) &  &  & \(R^2_\text{ch}(\text{fm}^2)\) & \(R^4_\text{ch}(\text{fm}^4)\) & \(R^6_\text{ch}(\text{fm}^6)\) & \(R^8_\text{ch}(\text{fm}^8)\)\\
             \hline
             $^{17}\text{Ne}$& 9.2495 & 8.9943 & 135.1084 & 3173.9365 & 121286.8172 &  & $^{26}\text{P}$& 10.4226 & 175.8714 & 4692.2968 & 216991.2120\\
             $^{19}\text{Ne}$& 9.0493 & 8.5873 & 114.3399 & 2147.7703 & 56244.9341 &  & $^{27}\text{P}$& 10.1694 & 163.6183 & 3980.0631 & 156415.3114\\
             $^{21}\text{Ne}$& 8.8179 & 8.3710 & 105.6433 & 1809.8420 & 40860.4523 &  & $^{28}\text{P}$& 9.8566 & 151.6761 & 3451.1493 & 122564.7524\\
             $^{23}\text{Ne}$& 8.4704 & 8.2087 & 99.5688 & 1590.9169 & 31806.3044 &  & $^{29}\text{P}$& 9.6290 & 142.9983 & 3074.9932 & 99658.9464\\
             $^{25}\text{Ne}$& 8.5943 & 8.1574 & 97.1154 & 1480.9113 & 26291.2946 &  & $^{30}\text{P}$& 9.7099 & 143.0606 & 2933.5918 & 83949.1405\\
             $^{27}\text{Ne}$& & 8.4336 & 103.0788 & 1577.5275 & 26072.5822 &  & $^{31}\text{P}$& 9.7817 & 143.6849 & 2865.9928 & 76276.5076\\
             $^{29}\text{Ne}$& & 8.8540 & 111.9464 & 1742.2310 & 28198.9593 &  & $^{33}\text{P}$& 10.1121 & 150.2262 & 2951.6227 & 74580.9270\\
             $^{31}\text{Ne}$& & 9.0500 & 113.2962 & 1042.1944 & -183910.9422 &  & $^{35}\text{P}$& 10.4318 & 156.9317 & 3060.3389 & 74664.5458\\
        \end{tabular}
    \end{ruledtabular}
\end{table*}

Ref. \cite{plb775126} reports that the charge density distribution of halo nuclei exhibits a more diffuse tail. Motivated by this, the nuclear charge density distributions for the Ne and P isotopic chains were calculated using the RCHB theory, incorporating intrinsic electromagnetic structure corrections. In this approach, the center-of-mass correction is applied only to the energy, and unlike in the treatment of Ref. \cite{prc034319}, the center-of-mass correction is not explicitly included in the calculation of nuclear radii in the present work. As illustrated in Fig. \ref{fig:1}, the charge density distributions of the confirmed halo nuclei $^{31}\text{Ne}$ and $^{17}\text{Ne}$ \cite{plb707357, prl262501, npa635292, plb57121} clearly exhibit a more diffuse tail compared to those of other isotopes. A similar diffuseness is observed for the neutron halo candidate $^{29}\text{Ne}$ \cite{plb707357} and the proton halo candidates $^{26,27,28}\text{P}$ \cite{prl815089}, supporting their proposed halo structures. Furthermore, as indicated by the dashed line in Fig. \ref{fig:1(a)}, negative charge densities emerge near the nuclear surface for many neutron-rich isotopes along the Ne chain. This phenomenon originates from the contribution of the neutron charge density distribution, which is incorporated via the intrinsic electromagnetic structure corrections. Since the neutron charge density is negative in the surface region \cite{prl112001,prl232002,nc121759}, its accumulation from a large number of excess neutrons in these neutron-rich systems results in negative values in the total nuclear charge density.

While the point-proton density distribution serves as a conventional approximation for the nuclear charge density, it constitutes an oversimplification. This is notably evident in the case of neutron-rich halo nuclei, where the contributions from neutron charge distributions become significant. A more accurate description requires the inclusion of contributions from the finite sizes of nucleons and their spin-orbit couplings, which are known to significantly affect both the charge density in halo nuclei~\cite{prc014320} and the charge radius \cite{prc045503}.
To illustrate the difference between the point-proton and charge densities, two confirmed halo nuclei and two candidate halo nuclei are compared in Fig. \ref{fig:2}. As shown in Figs. \ref{fig:2(a)} and \ref{fig:2(c)}, the charge density closely follows the point-proton density in proton-halo nuclei. In contrast, a notable deviation emerges in the tail region for neutron-halo nuclei. This difference can be attributed to the properties of the neutron electromagnetic form factor: the neutron itself possesses a negative charge density in its surface region \cite{prl112001}. Consequently, the accumulation of this negative charge from numerous neutrons in a diffuse neutron halo leads to the observed suppression in the total charge density. Similarly, the same phenomenon is observed in the halo nuclei candidates $^{26}\text{P}$ and $^{29}\text{Ne}$, as shown in Figs. \ref{fig:2(c)} and \ref{fig:2(d)}. This consistency is further complemented by the point-neutron density distributions provided in Fig. \ref{fig:2}, which serve as a crucial reference for identifying halo structures.\par

The charge form factor provides a momentum-space representation of the charge density and is more directly accessible experimentally than the spatial density distribution. Therefore, to investigate the halo structure extending to the nuclear surface, the behavior of the form factor at low momentum transfer is examined.\par

The squared differences of the charge form factors from the reference nuclei \(^{20}\mathrm{Ne}\)  and \(^{31}\mathrm{P}\) were calculated for the Ne and P isotopic chains. As shown in Figs.~\ref{fig:3(a)}, the known halo nuclei \(^{31,17}\mathrm{Ne}\) and candidate one \(^{29}\mathrm{Ne}\) along the neon isotopic chain all exhibit the most pronounced deviations. This suggests that a significant deviation of the charge form factor can serve as a sensitive probe for halo structures. \par

 As shown in Fig.~\ref{fig:3(b)}, the phosphorus isotopic chain presents a more complex picture: while candidate halo nuclei \(^{26}\mathrm{P}\) and \(^{27}\mathrm{P}\) show clear deviations, the non-halo nucleus \(^{35}\mathrm{P}\) exhibits an even larger one, whereas another candidate \(^{28}\mathrm{P}\) shows no significant deviation. This highlights the limitation of using a large charge form factor deviation alone as a halo criterion. Notably, the charge form factor in the low momentum region is also highly sensitive to the charge radius. Thus, the pronounced deviation of \(^{35}\mathrm{P}\) mainly arises from its larger charge radius (Table \ref{tab:table2}). Moreover, Fig.~\ref{fig:1(b)} shows that this radius increase stems from higher internal charge density around \(4~\text{fm}\), not from a diffuse surface tail. Therefore, a significant form factor deviation should be combined with an anomalous charge radius increase or specific density distribution features to reliably signal a diffuse halo. \par

Figure \ref{fig:3(c)} compares the squared charge and point-proton form factors for \(^{17}\mathrm{Ne}\), \(^{31}\mathrm{Ne}\), \(^{26}\mathrm{P}\), and \(^{29}\mathrm{Ne}\), revealing a clear contrast between proton and neutron halo nuclei at low momentum. The more pronounced difference in proton halos is attributed to their more diffuse proton distribution. Furthermore, the greater difference observed in \(^{31}\mathrm{Ne}\) compared to \(^{29}\mathrm{Ne}\) is consistent with its more diffuse charge density, as shown in Fig. \ref{fig:1(a)}.\par

The charge radius serves as a key identifier for proton halo nuclei. The valence protons directly induce a substantial enhancement, contrasting with the indirect and subtle increase in neutron halos caused by effects like the center-of-mass motion of the core within the halo. Consequently, proton halos typically exhibit significantly larger charge radii than their immediate neighbors \cite{plb775126}. Based on this, the charge radii for halo and candidate nuclei in the Ne and P isotopic chains were calculated using Eq. \eqref{eq:6}, which are presented in Table \ref{tab:table1}. Meanwhile, Fig. \ref{fig:o} presents a comparison of the experimental charge radii \cite{adndt9969} for the Ne isotopic chain with the calculated charge radii obtained from the corrected charge densities and the point-proton radii derived from the point-proton densities. Taking the Ne isotopes as an example, the charge radius of $^{17}\text{Ne}$ is even larger than that of $^{29}\mathrm{Ne}$ and close to $^{31}\text{Ne}$, clearly reflecting the diffuse proton distribution characteristic of proton halo nuclei. A consistent trend is observed in the point-proton radius , and it can be seen that the charge radii incorporating the intrinsic electromagnetic structure corrections agree more favorably with the experimental data than those obtained from the point-proton densities, confirming the necessity of incorporating these corrections. Meanwhile, the point-neutron radius of $^{31}\mathrm{Ne}$ is significantly larger than that of $^{29}\mathrm{Ne}$, consistent with the distribution features of neutron halo nuclei. For the P isotopes, the charge radii of $^{26}\mathrm{P}$, $^{27}\mathrm{P}$, and $^{28}\mathrm{P}$ show a monotonic decrease. The anomalously large charge radii of $^{26}\mathrm{P}$ and $^{27}\mathrm{P}$, combined with the extended tail in their charge density distributions (Figs. \ref{fig:1(b)} and \ref{fig:2(c)}), suggest possible proton halo structures. In contrast, the charge form factor of $^{28}\mathrm{P}$ in the low-momentum region (Fig. \ref{fig:3(b)}) and its relatively small charge radius show no clear signatures of a halo nucleus. All calculated values, along with individual electromagnetic structure corrections, are summarized in Table~\ref{tab:table1}, with the proton charge radius and mean square charge radius of the neutron taken from Refs.~\cite{RMP025010,nc121759}.\par

As defined in Eq. (\ref{eq:5}), the higher-order radial moment of the charge density effectively acts as a distance-weighted measure of the distribution. Consequently, at higher orders, these moments exhibit heightened sensitivity to the charge density at the nuclear surface. The eighth-order charge radial moments are calculated and compared for the F, Ne, Mg, and Ca isotopic chains in Fig. \ref{fig:4}. Pronounced oscillations are observed in regions of confirmed neutron halo nuclei as well as across most predicted ones. In the figure, the blue curve represents the point-proton density, while the red curve includes intrinsic electromagnetic structure corrections, where these oscillations become particularly evident. As the eighth-order moment involves the tenth power [cf. Eq. (\ref{eq:5})] of the radial coordinate, it amplifies the contribution from the tail of the density distribution and thus may be more sensitive to the choice of the box size. However, our further calculations show that varying the box size only affects the magnitude, without changing the oscillations.The details are also provided in the Appendix, including the oscillations under different box sizes and a specific discussion of the numerical variations.\par
An interesting observation is the initial appearance of oscillations at \(^{35}\mathrm{Mg}\) in the Mg chain (Fig.~\ref{fig:4(b)}), a nucleus not typically classified as a halo nucleus \cite{plb138422}. This may be associated with significant deformation occurring on the neutron-rich side of the Mg isotopes. It is important to note that the present RCHB calculations may have limitations in describing halo phenomena in deformed nuclei. To address this, our future work will employ the DRHBc theory to investigate the charge properties of halo nuclei. The fourth-order charge radial moments have also been calculated for the F, Ne, Mg, and Ca isotopic chains. Although oscillations appear in the regions of established and predicted neutron halo nuclei, they are less pronounced than those in the eighth-order moments. The corresponding higher-order moments for the Ne and P chains are provided in Table \ref{tab:table2}, where the available experimental charge radii of Ne isotopes are also included as a benchmark.

\section{\label{sec:4}SUMMARY AND PROSPECTS}

\begin{figure*}
\centering
\includegraphics[width=16cm]{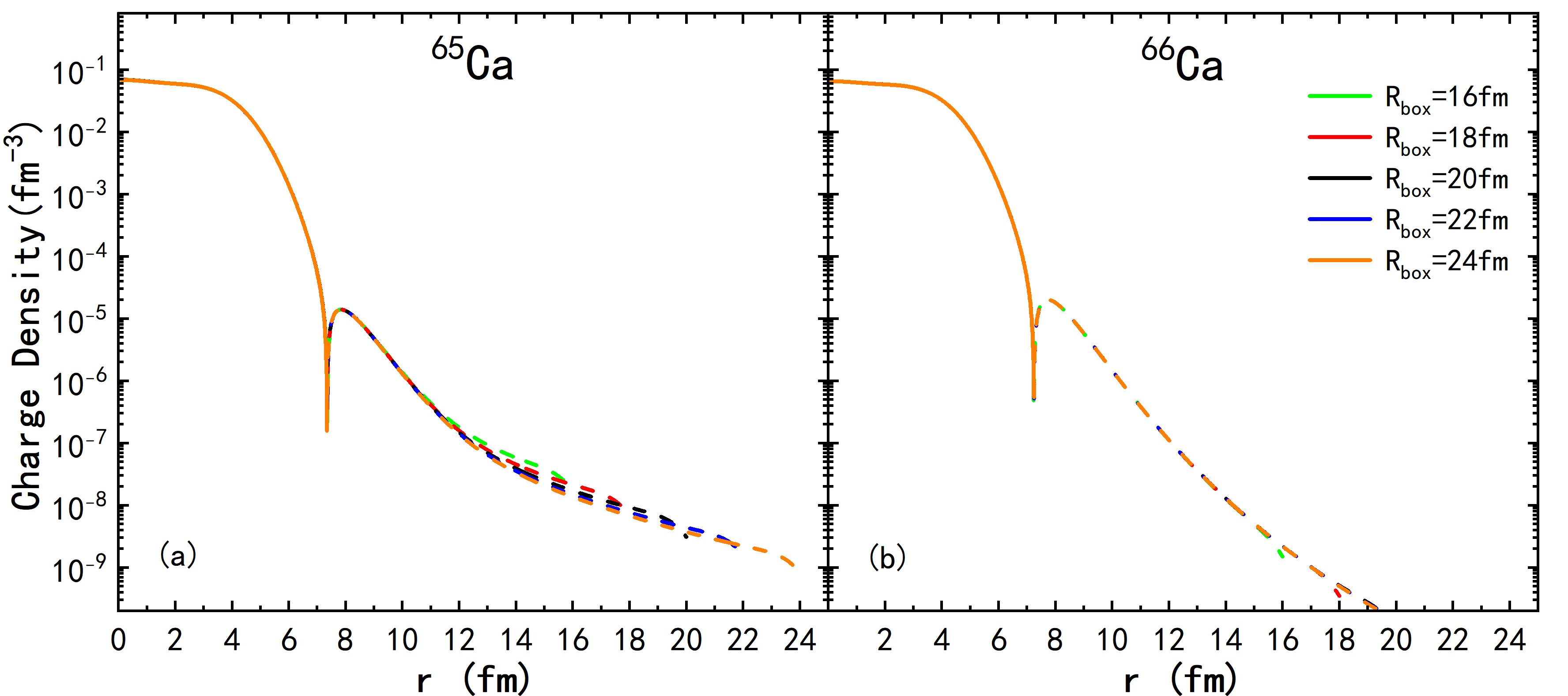}
\caption{\raggedright\label{fig:wide}Charge density distributions for $^{65}\text{Ca}$ and $^{66}\text{Ca}$ with different box sizes, obtained from the RCHB theory with intrinsic electromagnetic structure corrections, with dashed lines indicating the absolute values of the negative charge densities.
}\label{appendixfig:1}
   \subcaptionbox{\label{appendixfig:1(a)}}{}
   \subcaptionbox{\label{appendixfig:1(b)}}{}
\end{figure*}

This study has conducted an investigation into the charge properties of halo nuclei,  with a primary focus on the Ne and P isotopic chains using RCHB theory. A key aspect of the approach has been the incorporation of corrections for the intrinsic electromagnetic structure, specifically accounting for the finite size of nucleons and the contributions from the spin-orbit density.\par

The results reveal distinctive characteristics of halo nuclei. Their charge density distributions are marked by an extended tail, and a clear contrast is observed between the profiles of proton halos and neutron halos when compared to their respective point-proton densities. Analysis of the charge form factors further underscores these unique features, showing pronounced deviations from those of reference nuclei at low momentum transfer. Significant differences were also identified between the charge and point-proton form factors for both types of halos. Notably, halo nuclei are also characterized by anomalously large charge radii or point-nucleon radii compared to their neighboring isotopes. Moreover, the eighth-order moments of the charge density exhibit abrupt changes and pronounced oscillations in the confirmed neutron halo nuclei and most predicted ones along the Ne, Mg, F, and Ca isotopic chains. These oscillations serve as a clear signature of substantial structural modifications occurring on the nuclear charge density surface.\par

Prospectively, with the advancement of experimental techniques, particularly the development of technologies such as self-confining RI ion target (SCRIT), direct and precise measurements of the charge density distributions of unstable nuclei have become possible \cite{prl092502}. The systematic features revealed in this work offer theoretical references and observable physical signatures for future detection and identification of halo nuclei using electron scattering and other methods.\par

While the present work is based on the spherical RCHB theory, future research will expand upon these findings by employing the DRHBc theory with the self-consistent intrinsic electromagnetic structure
correction. 
The DRHBc framework has already been widely applied to describe deformed nuclei such as Ne isotopes \cite{plb138792}. Building on this, our future work will further extend this approach by incorporating the intrinsic electromagnetic structure corrections considered in the present study, aiming to systematically explore the charge properties of deformed halo nuclei. Such investigations are expected to contribute to a more comprehensive understanding of nuclear structure in these exotic systems.

\begin{acknowledgments}
This work is supported by the National Natural Science Foundation of China (Grant No.12475119, No.12447101 and No.12305125) and the National Key Laboratory of Neutron Science and Technology (Grant No. NST202401016). The authors would like to express their sincere appreciation to Dr. Xiao Lu for her helpful discussions and suggestions.
\end{acknowledgments}

\section*{\label{sec:appendix}APPENDIX: CONVERGENCE CHECKS FOR CHARGE DENSITY, RADII, AND HIGHER-ORDER MOMENTS}
\begin{figure}
\centering
\includegraphics[width=9cm]{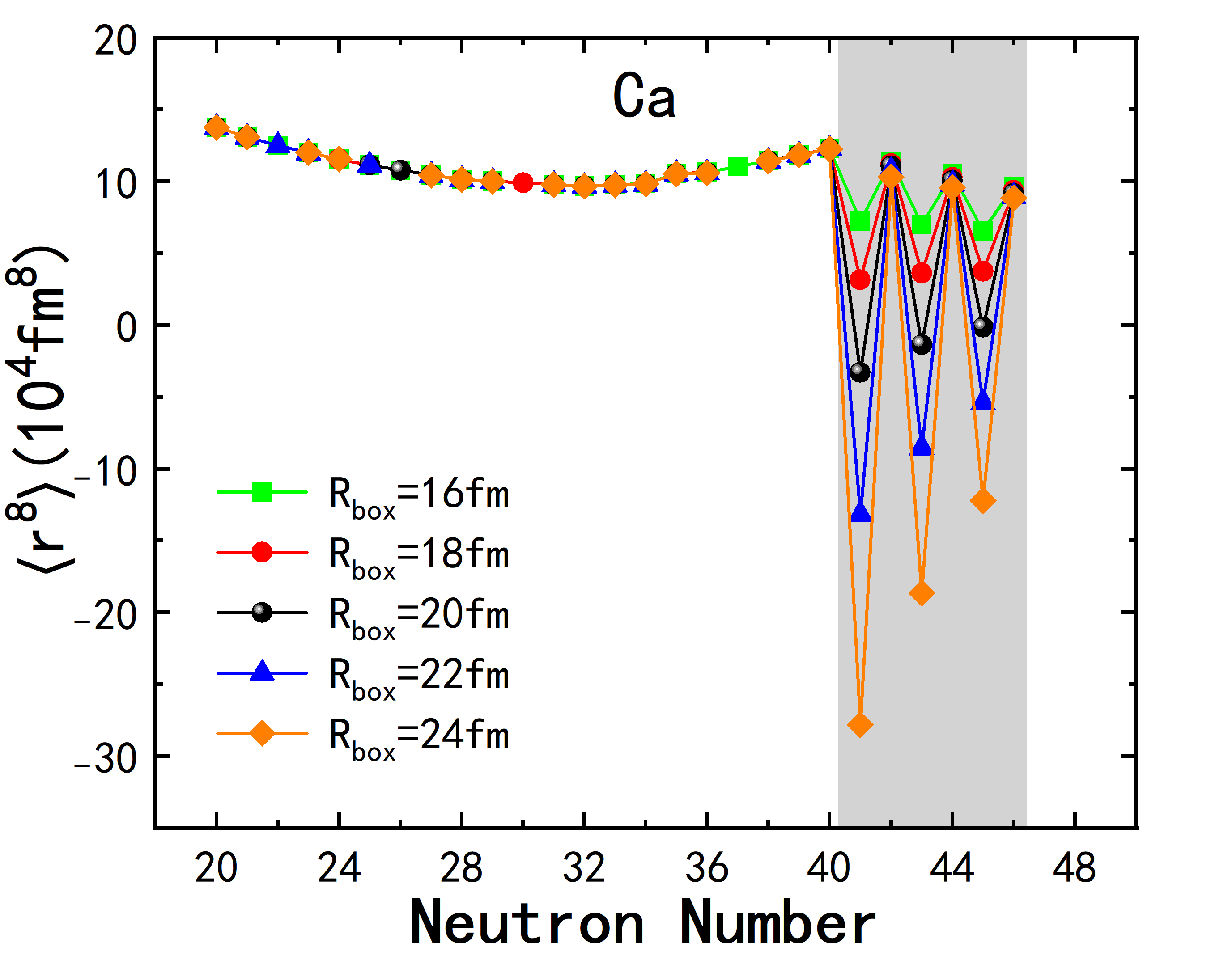}
\caption{\raggedright\label{fig:wide}Eighth-order charge radial moments for the Ca isotopic chain, calculated using different box sizes. The gray shaded areas denote nuclei predicted to exhibit a halo structure.}\label{appendixfig:2}
\end{figure}

To verify that a box size of 20 fm is reasonable for the present study, we calculated the charge density distributions of $^{65}\text{Ca}$ and $^{66}\text{Ca}$ with $R_\text{box}=16,18,20,22,\;\text{and}\;24\;\text{fm}$. The results are shown in Fig. \ref{appendixfig:1}. To avoid renormalizing the zero-range pairing strength with respect to the corresponding model space while maintaining the convergence of self-consistent iteration, we adopt a small pairing strength of $-150\;\text{MeV} \cdot \text{fm}^3$ for these checks, in which case the pairing energy is very small and remains nearly constant as $R_\text{box}$ varies. In the region $r<12\;\text{fm}$, the charge densities obtained with different box sizes nearly coincide, showing indistinguishable difference. For $r\geq 12\;\text{fm}$, the even–even nucleus $^{66}\text{Ca}$ still exhibits marginal differences. For the odd-$A$ nucleus $^{65}\text{Ca}$, some differences can be observed, but from $R_\text{box} =20 \;\text{fm}$ to 24 fm, the differences are very small, and the charge densities remain within the same order of magnitude. The greater sensitivity of $^{65}\text{Ca}$ to the box size, compared to $^{66}\text{Ca}$, originates from its more diffuse density distribution, which makes the tail more dependent on the box size. Nevertheless, for both $^{66}\text{Ca}$ and $^{65}\text{Ca}$, the negative density contribution at the level of $10^{-5}$ to $10^{-6}$ remains robust.\par

From Fig. \ref{appendixfig:1}, it is also evident that regardless of the box size, the charge density distribution of $^{65}\text{Ca}$ is consistently more diffuse than that of $^{66}\text{Ca}$. This supports the discussion on higher-order moments in this work, which will be elaborated in detail below.\par

\begin{table}[t] 
    \caption{\label{tab:table3}%
Charge radii and higher-order moments of the charge density for $^{65}\text{Ca}$ as functions of the box size.
    }
    \begin{ruledtabular}
        \begin{tabular}{ccccc}
             \(R_\text{box}\) & \(R_\text{ch}(\text{fm})\) & \(R^2_\text{ch}(\text{fm}^2)\) & \(R^4_\text{ch}(\text{fm}^4)\) & \(R^8_\text{ch}(\text{fm}^8)\) \\
             \hline
             16 & 3.68572 & 13.59568 & 247.90047 & 71100.37561 \\
             18 & 3.68555 & 13.56922 & 246.93686 & 42772.82539 \\
             20 & 3.68544 & 13.56783 & 246.62467 & 1520.77129 \\
             22 & 3.68535 & 13.56667 & 246.33312 & -56027.98020 \\
             24 & 3.68532 & 13.56565 & 246.05854 & -133519.58792 \\
             
        \end{tabular}
    \end{ruledtabular}
\end{table}

\begin{table}[t] 
    \caption{\label{tab:table4}%
Charge radii and higher-order moments of the charge density for $^{66}\text{Ca}$ as functions of the box size.
    }
    \begin{ruledtabular}
        \begin{tabular}{ccccc}
             \(R_\text{box}\) & \(R_\text{ch}(\text{fm})\) & \(R^2_\text{ch}(\text{fm}^2)\) & \(R^4_\text{ch}(\text{fm}^4)\) & \(R^8_\text{ch}(\text{fm}^8)\) \\
             \hline
             16 & 3.70348 & 13.71579 & 250.37508 & 96352.03068 \\
             18 & 3.70336 & 13.71486 & 250.32000 & 93578.99977 \\
             20 & 3.70333 & 13.71464 & 250.29788 & 91377.43087 \\
             22 & 3.70333 & 13.71462	& 250.28709	& 89613.81328 \\
             24 & 3.70334 & 13.71475 & 250.28414 & 88253.87567 \\
             
        \end{tabular}
    \end{ruledtabular}
\end{table}

We also calculated the charge radii and higher-order moments discussed in this work using the charge density distributions shown in Fig. \ref{appendixfig:1}. The results are presented in Table \ref{tab:table3} and \ref{tab:table4}. For $^{66}\text{Ca}$, when the box size increases from 20 fm to 24 fm, the charge radius changes by approximately 0.00001 fm, corresponding to a relative difference of only 0.00027$\%$. The second-order moment shows minor variations of the same order of magnitude. As the order increases, the calculation of higher-order moments effectively involves higher powers of the radial coordinate, which inevitably amplifies the contribution from the tail of the charge density, as illustrated by the Eq. (\ref{eq:5}). Nevertheless, for the fourth-order moment, the change is only 0.014 $\text{fm}^4$,a relative difference of 0.0055$\%$. For $^{65}\text{Ca}$, the tail of the charge density distribution is more diffuse, making the tail structure more sensitive to the box size. Consequently, the charge radius and higher-order moments of $^{65}\text{Ca}$ exhibit larger variations with $R_\text{box}$  than those of $^{66}\text{Ca}$. Even so, when the box size increases from 20 fm to 24 fm, the charge radius changes by only about 0.00012 fm, a relative difference of 0.0032$\%$, and the fourth-order moment changes by 0.566 $\text{fm}^4$, a relative difference of less than 0.23$\%$.\par

For the eighth-order moment, which involves a weighting factor of $r^{10}$, the result becomes very sensitive to the charge density near the box boundary. However, the discussion of the eighth-order moment in this work does not focus on its absolute value, but rather qualitatively on the oscillatory behavior in the region where possible halo nuclei emerge. To this end, we show in Fig. \ref{appendixfig:2} the evolution of the eighth-order moment of Ca isotopes as a function of neutron number for different $R_\text{box}$ values. It can be seen that although the amplitude of oscillation increases with increasing $R_\text{box}$, the oscillatory behavior itself remains robust. This unambiguously supports our qualitative discussion.

\nocite{*}
\FloatBarrier
\bibliography{apssamp}

@article{plb707357,
  title={Interaction cross sections for Ne isotopes towards the island of inversion and halo
structures of $^{29}\text{Ne}$ and $^{31}\text{Ne}$},
  author={M. Takechi and T. Ohtsubo and M. Fukuda and D. Nishimura and T. Kuboki and others },
  journal={Phys. Lett. B},
  volume={707},
  pages={357},
  year={2012},
  doi = {https://doi.org/10.1016/j.physletb.2011.12.028},
  url = {https://www.sciencedirect.com/science/article/pii/S0370269311014857?via%3Dihub},
  publisher={Elsevier}
}

@article{npa635292,
  title={Reaction cross sections in Si of light proton-halo candidates $^{12}\text{N}$ and $^{17}\text{Ne}$},
  author={R.E. Warner and H. Thirumurthy and J. Woodroffe and F.D. Becchetti and J.A. Brown and B.S. Davids and others},
  journal={Nucl. Phys. A},
  volume={635},
  pages={292},
  year={1998},
  doi = {https://doi.org/10.1016/S0375-9474(98)00163-8},
  url = {https://www.sciencedirect.com/science/article/pii/S0375947498001638?via%3Dihub},
  publisher={Elsevier}
}

@article{plb57121,
  title={Possibility of a two-proton halo in $^{17}\text{Ne}$},
  author={R. Kanungo and M. Chiba and S. Adhikari and D. Fang and N. Iwasa and K. Kimura and others },
  journal={Phys. Lett. B},
  volume={571},
  pages={21},
  year={2003},
  doi = {https://doi.org/10.1016/j.physletb.2003.07.050},
  url = {https://www.sciencedirect.com/science/article/pii/S0370269303011031?via%3Dihub},
  publisher={Elsevier}
}

@article{prl815089,
  title={Spectroscopy of Radioactive Beams from Single-Nucleon Knockout Reactions: Application to the $sd$ Shell Nuclei $^{25}\text{Al}$ and $^{26,27,28}\text{P}$},
  author={A. Navin and D. Bazin and B. A. Brown and B. Davids and G. Gervais and T. Glasmacher and others},
  journal={Phys. Rev. Lett.},
  volume={81},
  pages={5089},
  year={1998},
  doi = {https://doi.org/10.1103/PhysRevLett.81.5089},
  url = {https://journals.aps.org/prl/abstract/10.1103/PhysRevLett.81.5089},
  publisher={APS}
}

@article{prl242501,
  title={Observation of a $p$-Wave One-Neutron Halo Configuration in $^{37}\text{Mg}$},
  author={N. Kobayashi1 and T. Nakamura1 and Y. Kondo1 and J.A. Tostevin and Y. Utsuno and N. Aoi and others},
  journal={Phys. Rev. Lett.},
  volume={112},
  pages={242501},
  year={2014},
  doi = { https://doi.org/10.1103/PhysRevLett.112.242501},
  url = {https://journals.aps.org/prl/abstract/10.1103/PhysRevLett.112.242501},
  publisher={APS}
}

@article{prl222504,
  title={Two-Neutron Halo is Unveiled in $^{29}\text{F}$},
  author={S. Bagchi and R. Kanungo and Y.K. Tanaka and H. Geissel and P. Doornenbal and W. Horiuchi and others},
  journal={Phys. Rev. Lett.},
  volume={124},
  pages={222504},
  year={2020},
  doi = {https://doi.org/10.1103/PhysRevLett.124.222504},
  url = {https://journals.aps.org/prl/abstract/10.1103/PhysRevLett.124.222504},
  publisher={APS}
}

@article{prc014328,
  title={Exploring the halo character and dipole response in the dripline nucleus $^{31}\text{F}$},
  author = {G. Singh and Jagjit Singh and J. Casal and and L. Fortunato},
  journal={Phys. Rev. C},
  volume={105},
  pages={014328},
  year={2022},
  doi = {https://doi.org/10.1103/PhysRevC.105.014328},
  url = {https://journals.aps.org/prc/abstract/10.1103/PhysRevC.105.014328},
  publisher={APS}
}

@article{prc041302,
  title={Giant halo at the neutron drip line in Ca isotopes in relativistic continuum Hartree-Bogoliubov theory},
  author = {J. Meng and H. Toki and J. Y. Zeng and S. Q. Zhang and S.G. Zhou},
  journal={Phys. Rev. C},
  volume={65},
  pages={041302},
  year={2002},
  doi = {https://doi.org/10.1103/PhysRevC.65.041302},
  url = {https://journals.aps.org/prc/abstract/10.1103/PhysRevC.65.041302},
  publisher={APS}
}

@article{prc054303,
  title={Effects of the neutron spin-orbit density on the nuclear charge density in relativistic models},
  author = {Haruki Kurasawa and Toshio Suzuki},
  journal={Phys. Rev. C},
  volume={62},
  pages={054303},
  year={2000},
  doi = {https://doi.org/10.1103/PhysRevC.62.054303},
  url = {https://journals.aps.org/prc/abstract/10.1103/PhysRevC.62.054303},
  publisher={APS}
}

@article{PTEP86127,
  title={The $n$th-order moment of the nuclear charge density and contribution from the neutrons},
  author={Haruki Kurasawa and Toshio Suzuki},
  journal={Prog. Theor. Exp. Phys.},
  volume={2019},
  pages={113D01},
  year={2019},
  doi = {https://doi.org/10.1093/ptep/ptz121},
  url = {https://academic.oup.com/ptep/article/2019/11/113D01/5645109},
  publisher={PTEP}
}

@article{prc054319,
  title={New parametrization for the nuclear covariant energy density functional with a point-coupling interaction},
  author = {P. W. Zhao and Z. P. Li and J. M. Yao  and J. Meng},
  journal={Phys. Rev. C},
  volume={82},
  pages={054319},
  year={2010},
  doi = {https://doi.org/10.1103/PhysRevC.82.054319},
  url = {https://journals.aps.org/prc/abstract/10.1103/PhysRevC.82.054319},
  publisher={APS}
}

@article{adndt1211,
  title={The limits of the nuclear landscape explored by the relativistic continuum Hartree–Bogoliubov theory},
  author={X.W. Xia and Y. Lim and P.W. Zhao and H.Z. Liang f and X.Y. Qu and others},
  journal={At. Data Nucl. Data Tables},
  volume={121},
  pages={1},
  year={2018},
  doi = {https://doi.org/10.1016/j.adt.2017.09.001},
  url = {https://www.sciencedirect.com/science/article/pii/S0092640X17300451?via%3Dihub},
  publisher={ELSEVIER}
}

@article{prA042807,
  title={Finite-nuclear-size effect in hydrogenlike ions with relativistic nuclear structure},
  author = {Hui Hui Xie and Jian Li and Li Guang Jiao and Yew Kam Ho},
  journal={Phys. Rev. A},
  volume={107},
  pages={042807},
  year={2023},
  doi = {https://doi.org/10.1103/PhysRevA.107.042807},
  url = {https://journals.aps.org/pra/abstract/10.1103/PhysRevA.107.042807},
  publisher={APS}
}

@article{PPNP59694,
  title={Nucleon electromagnetic form factors},
  author={C.F. Perdrisat and V. Punjabi and M. Vanderhaeghen},
  journal={Prog. Part. Nucl. Phys.},
  volume={59},
  pages={694},
  year={2007},
  doi = {https://doi.org/10.1016/j.ppnp.2007.05.001},
  url = {https://www.sciencedirect.com/science/article/pii/S0146641007000610},
  publisher={ELSEVIER}
}

@article{plb775126,
  title={Two-proton radioactivity with 2p halo in light mass nuclei $\textit{A} = 18–34$},
  author={G. Saxena and M. Kumawat and M. Kaushik and S.K. Jain and Mamta Aggarwal },
  journal={Phys. Lett. B},
  volume={775},
  pages={126},
  year={2017},
  doi = {https://doi.org/10.1016/j.physletb.2017.10.055},
  url = {https://www.sciencedirect.com/science/article/pii/S0370269317308675},
  publisher={Elsevier}
}

@article{prc014320,
  title={Effect of spin-orbit nuclear charge density corrections due to the anomalous magnetic moment on halonuclei},
  author = {A. Ong and J. C. Berengut and V. V. Flambaum},
  journal={Phys. Rev. C},
  volume={82},
  pages={014320},
  year={2010},
  doi = {https://doi.org/10.1103/PhysRevC.82.014320},
  url = {https://journals.aps.org/prc/abstract/10.1103/PhysRevC.82.014320},
  publisher={APS}
}

@article{prc045503,
  title={Impact of spin-orbit currents on the electroweak skin of neutron-rich nuclei},
  author = {C. J. Horowitz and J. Piekarewicz},
  journal={Phys. Rev. C},
  volume={86},
  pages={045503},
  year={2012},
  doi = {https://doi.org/10.1103/PhysRevC.86.045503},
  url = {https://journals.aps.org/prc/abstract/10.1103/PhysRevC.86.045503},
  publisher={APS}
}

@article{prl112001,
  title={Charge Densities of the Neutron and Proton},
  author={Gerald A. Miller},
  journal={Phys. Rev. Lett.},
  volume={99},
  pages={112001},
  year={2007},
  doi = {https://doi.org/10.1103/PhysRevLett.99.112001},
  url = {https://journals.aps.org/prl/abstract/10.1103/PhysRevLett.99.112001},
  publisher={APS}
}

@article{RMP025010,
  title={CODATA recommended values of the fundamental physical constants: 2018},
  author={Eite Tiesinga and Peter J. Mohr and David B. Newell and Barry N. Taylor},
  journal={Rev. Mod. Phys.},
  volume={93},
  pages={025010},
  year={2021},
  doi = {https://doi.org/10.1103/RevModPhys.93.025010},
  url = {https://journals.aps.org/rmp/abstract/10.1103/RevModPhys.93.025010},
  publisher={APS}
}

@article{nc121759,
  title={Measurement of the neutron charge radius and the role of its constituents},
  author={H. Atac and M. Constantinou and Z.E. Meziani and M. Paolone and N. Sparveris},
  journal={Nat. Commun.},
  volume={12},
  pages={1759},
  year={2021},
  doi = {https://doi.org/10.1038/s41467-021-22028-z},
  url = {https://www.nature.com/articles/s41467-021-22028-z},
  publisher={nature}
}

@article{plb138422,
  title={A unified description of the halo nucleus $^{37}\text{Mg}$ from microscopic structure to reaction observables},
  author={Jia Lin An and Kai Yuan Zhang and Qi Lu and Shi Yi Zhong and Shi Sheng Zhang },
  journal={Phys. Lett. B},
  volume={849},
  pages={138422},
  year={2024},
  doi = {https://doi.org/10.1016/j.physletb.2023.138422},
  url = {https://www.sciencedirect.com/science/article/pii/S0370269323007554},
  publisher={Elsevier}
}

@article{RMP76215,
  title={Structure and reactions of quantum halos},
  author={A. S. Jensen and K. Riisager and D. V. Fedorov and E. Garrido},
  journal={Rev. Mod. Phys.},
  volume={76},
  pages={215},
  year={2004},
  doi = {https://doi.org/10.1103/RevModPhys.76.215},
  url = {https://journals.aps.org/rmp/abstract/10.1103/RevModPhys.76.215},
  publisher={APS}
}

@article{prl80460,
  title={Giant Halo at the Neutron Drip Line},
  author={J. Meng and P. Ring},
  journal={Phys. Rev. Lett.},
  volume={80},
  pages={460},
  year={1998},
  doi = {https://doi.org/10.1103/PhysRevLett.80.460},
  url = {https://journals.aps.org/prl/abstract/10.1103/PhysRevLett.80.460},
  publisher={APS}
}

@article{PPNP68215,
  title={Recent experimental progress in nuclear halo structure studies},
  author={Isao Tanihata and Herve Savajols and Rituparna Kanungo },
  journal={Prog. Part. Nucl. Phys.},
  volume={68},
  pages={215},
  year={2013},
  doi = {https://doi.org/10.1016/j.ppnp.2012.07.001},
  url = {https://www.sciencedirect.com/science/article/pii/S0146641012001081},
  publisher={ELSEVIER}
}

@article{prl222501,
  title={Nuclear Structure of Dripline Nuclei Elucidated through Precision Mass Measurements of $^{23}\text{Si}$, $^{26}\text{P}$, $^{27,28}\text{S}$, and $^{31}\text{Ar}$},
  author={Y. Yu and Y.M. Xing and Y.H. Zhang and M. Wang and X.H. Zhou and J.G. Li and others},
  journal={Phys. Rev. Lett.},
  volume={133},
  pages={222501},
  year={2024},
  doi = {https://doi.org/10.1103/PhysRevLett.133.222501},
  url = {https://journals.aps.org/prl/abstract/10.1103/PhysRevLett.133.222501},
  publisher={APS}
}

@article{prc014325,
  title={Effective ﬁeld theory for proton halo nuclei},
  author = {Emil Ryberg and Christian Forsse´n and H.W. Hammer and Lucas Platter},
  journal={Phys. Rev. C},
  volume={89},
  pages={014325},
  year={2014},
  doi = {https://doi.org/10.1103/PhysRevC.89.014325},
  url = {https://journals.aps.org/prc/abstract/10.1103/PhysRevC.89.014325},
  publisher={APS}
}

@article{prc044325,
  title={Examining possible neutron-halo nuclei heavier than $^{37}\text{Mg}$},
  author = {Ikuko Hamamoto},
  journal={Phys. Rev. C},
  volume={95},
  pages={044325},
  year={2017},
  doi = {https://doi.org/10.1103/PhysRevC.95.044325},
  url = {https://journals.aps.org/prc/abstract/10.1103/PhysRevC.95.044325},
  publisher={APS}
}

@article{prc064313,
  title={Structure of $^{22}\text{N}$ and the $\textit{N}=14$ subshell},
  author = {C. Rodr{\'i}guez-Tajes and D. Cortina Gil and H. {\'A}lvarez-Pol and T. Aumann and E. Benjamim and others},
  journal={Phys. Rev. C},
  volume={83},
  pages={064313},
  year={2011},
  doi = {https://doi.org/10.1103/PhysRevC.83.064313},
  url = {https://journals.aps.org/prc/abstract/10.1103/PhysRevC.83.064313},
  publisher={APS}
}

@article{prc0643132,
  title={Nuclear interactions in weakly bound neutron-rich nuclei},
  author = {B. Mukeru and M. B. Mahatikele and G. J. Rampho},
  journal={Phys. Rev. C},
  volume={107},
  pages={064313},
  year={2023},
  doi = {https://doi.org/10.1103/PhysRevC.107.064313},
  url = {https://journals.aps.org/prc/abstract/10.1103/PhysRevC.107.064313},
  publisher={APS}
}

@article{prc024320,
  title={Research on exotic nuclei in deformed relativistic mean-field theory plus BCS in complex momentum representation},
  author = {Yu Xuan Luo and Quan Liu and Jian You Guo},
  journal={Phys. Rev. C},
  volume={108},
  pages={024320},
  year={2023},
  doi = {https://doi.org/10.1103/PhysRevC.108.024320},
  url = {https://journals.aps.org/prc/abstract/10.1103/PhysRevC.108.024320},
  publisher={APS}
}

@article{npa703202,
  title={First-forbidden mirror $\beta$-decays in $\textit{A}=17$ mass region},
  author={N. Michel and J. Okołowicz and F. Nowacki and M. Płoszajczak},
  journal={Nucl. Phys. A},
  volume={703},
  pages={212},
  year={2002},
  doi = {https://doi.org/10.1016/S0375-9474(01)01529-9},
  url = {https://www.sciencedirect.com/science/article/pii/S0375947401015299},
  publisher={ELSEVIER}
}

@article{plb69790,
  title={Constraints on two-neutron separation energy in the Borromean $^{22}\text{C}$ nucleus},
  author={M.T. Yamashita and R.S. Marques de Carvalho and T. Frederico and Lauro Tomio },
  journal={Phys. Lett. B},
  volume={697},
  pages={90},
  year={2011},
  doi = {https://doi.org/10.1016/j.physletb.2011.01.040},
  url = {https://www.sciencedirect.com/science/article/pii/S0370269311000773},
  publisher={Elsevier}
}

@article{prc057301,
  title={Charge and matter radii of Borromean halo nuclei: The $^{6}\text{He}$ nucleus},
  author = {B. V. Danilin and S. N. Ershov and J. S. Vaagen},
  journal={Phys. Rev. C},
  volume={71},
  pages={057301},
  year={2005},
  doi = {https://doi.org/10.1103/PhysRevC.71.057301},
  url = {https://journals.aps.org/prc/abstract/10.1103/PhysRevC.71.057301},
  publisher={APS}
}

@article{prc024318,
  title={Charge and matter form factors of two-neutron halo nuclei in halo effective field theory at next-to-leading order},
  author = {Jared Vanasse},
  journal={Phys. Rev. C},
  volume={95},
  pages={024318},
  year={2017},
  doi = {https://doi.org/10.1103/PhysRevC.95.024318},
  url = {https://journals.aps.org/prc/abstract/10.1103/PhysRevC.95.024318},
  publisher={APS}
}

@article{prc014610,
  title={Charge density distributions and charge form factors of the $\textit{N}=82$ and $\textit{N}=126$ isotonic nuclei},
  author={Zaijun Wang and Zhongzhou Ren and Ying Fan},
  journal={Phys, Rev. C.},
  volume={73},
  pages={014610},
  year={2006},
  doi = {https://doi.org/10.1103/PhysRevC.73.014610},
  url = {https://journals.aps.org/prc/abstract/10.1103/PhysRevC.73.014610},
  publisher={APS}
}

@article{prc044307,
  title={Charge and matter distributions and form factors of light, medium, and heavy neutron-rich nuclei},
  author={A. N. Antonov and D. N. Kadrev and M. K. Gaidarov and E. Moya de Guerra and P. Sarriguren and others},
  journal={Phys, Rev. C.},
  volume={72},
  pages={044307},
  year={2005},
  doi = {https://doi.org/10.1103/PhysRevC.72.044307},
  url = {https://journals.aps.org/prc/abstract/10.1103/PhysRevC.72.044307},
  publisher={APS}
}

@article{prcL041303,
  title={Collapse of the $\textit{N}=28$ shell closure in the newly discovered $^{39}\text{Na}$ nucleus and the development of deformed halos towards the neutron dripline},
  author = {K. Y. Zhang and P. Papakonstantinou and M.H. Mun and Y. Kim and H. Yan and X.X. Sun},
  journal={Phys. Rev. C},
  volume={107},
  pages={L041303},
  year={2023},
  doi = {https://doi.org/10.1103/PhysRevC.107.L041303},
  url = {https://journals.aps.org/prc/abstract/10.1103/PhysRevC.107.L041303},
  publisher={APS}
}

@article{prc0143202,
  title={Examination of the evidence for a proton halo in $^{22}\text{Al}$},
  author = {K. Y. Zhang and C. Pan and Si Bo Wang},
  journal={Phys. Rev. C},
  volume={110},
  pages={014320},
  year={2024},
  doi = {https://doi.org/10.1103/PhysRevC.110.014320},
  url = {https://journals.aps.org/prc/abstract/10.1103/PhysRevC.110.014320},
  publisher={APS}
}

@article{npa422103,
  title={Hartree-Fock-Bogolyubov description of nuclei near the neutron-drip line},
  author={J. Dobaczewski and H. Flocard and J. Treiner},
  journal={Nucl. Phys. A},
  volume={422},
  pages={103},
  year={1984},
  doi = {https://doi.org/10.1016/0375-9474(84)90433-0},
  url = {https://www.sciencedirect.com/science/article/abs/pii/0375947484904330},
  publisher={ELSEVIER}
}

@article{npa6353,
  title={Relativistic continuum Hartree-Bogoliubov theory with both zero range and finite range Gogny force and their application},
  author={Jie Meng},
  journal={Nucl. Phys. A},
  volume={635},
  pages={3},
  year={1998},
  doi = {https://doi.org/10.1016/S0375-9474(98)00178-X},
  url = {https://www.sciencedirect.com/science/article/pii/S037594749800178X},
  publisher={ELSEVIER}
}

@article{prl232002 ,
  title={Charge Distributions of Moving Nucleons},
  author={Ce´dric Lorce},
  journal={Phys. Rev. Lett.},
  volume={125},
  pages={232002},
  year={2020},
  doi = {https://doi.org/10.1103/PhysRevLett.125.232002},
  url = {https://journals.aps.org/prl/abstract/10.1103/PhysRevLett.125.232002},
  publisher={APS}
}

@article{prl262501 ,
  title={Halo Structure of the Island of Inversion Nucleus $^{31}\text{Ne}$},
  author={T. Nakamura and N. Kobayashi and Y. Kondo and Y. Satou and N. Aoi and others},
  journal={Phys. Rev. Lett.},
  volume={103},
  pages={262501},
  year={2009},
  doi = {https://doi.org/10.1103/PhysRevLett.103.262501},
  url = {https://journals.aps.org/prl/abstract/10.1103/PhysRevLett.103.262501},
  publisher={APS}
}

@article{prl142501 ,
  title={Deformation-Driven p-Wave Halos at the Drip Line:  $^{31}\text{Ne}$},
  author={T. Nakamura and N. Kobayashi and Y. Kondo and Y. Satou and others},
  journal={Phys. Rev. Lett.},
  volume={112},
  pages={142501},
  year={2014},
  doi = {https://doi.org/10.1103/PhysRevLett.112.142501},
  url = {https://journals.aps.org/prl/abstract/10.1103/PhysRevLett.112.142501},
  publisher={APS}
}

@article{prc014312,
  title={Constraints on the neutron drip line with the newly observed $^{39}\text{Na}$},
  author = {Q. Z. Chai and J. C. Pei and Na Fei and D. W. Guan},
  journal={Phys. Rev. C},
  volume={102},
  pages={014312},
  year={2020},
  doi = {https://doi.org/10.1103/PhysRevC.102.014312},
  url = {https://journals.aps.org/prc/abstract/10.1103/PhysRevC.102.014312},
  publisher={APS}
}

@article{plb139082,
  title={Charge radii of $^{11-16}\text{C}$, $^{13-17}\text{N}$ and $^{15-18}\text{O}$ determined from their charge-changing cross-sections and the mirror-difference charge radii},
  author={J.W. Zhao and B.H. Sun and I. Tanihata and J.Y. Xu and others  },
  journal={Phys. Lett. B},
  volume={858},
  pages={139082},
  year={2024},
  doi = {https://doi.org/10.1016/j.physletb.2024.139082},
  url = {https://www.sciencedirect.com/science/article/pii/S0370269324006403},
  publisher={Elsevier}
}

@article{prc014001,
  title={Effective field theory for weakly bound two-neutron halo nuclei: Corrections from neutron-neutron effective range},
  author = {Davi B. Costa and Masaru Hongo and Dam Thanh Son},
  journal={Phys. Rev. C},
  volume={112},
  pages={014001},
  year={2025},
  doi = {https://doi.org/10.1103/lds3-g3tp},
  url = {https://doi.org/10.1103/lds3-g3tp},
  publisher={APS}
}

@article{prc014306,
  title={High-precision ab initio radius calculations of boron isotopes},
  author = {T. Wolfgruber and T. Gesser and M. Knöll and P. Maris and R. Roth},
  journal={Phys. Rev. C},
  volume={112},
  pages={014306},
  year={2025},
  doi = {https://doi.org/10.1103/8mfb-wc36},
  url = {https://journals.aps.org/prc/abstract/10.1103/8mfb-wc36},
  publisher={APS}
}

@article{prl162503 ,
  title={Ab Initio Study of the Beryllium Isotopes $^{7}\text{Be}$ to $^{12}\text{Be}$},
  author={Shihang Shen and Serdar Elhatisari and Dean Lee and Ulf G. Meißner and Zhengxue Ren},
  journal={Phys. Rev. Lett.},
  volume={134},
  pages={162503},
  year={2025},
  doi = {https://doi.org/10.1103/PhysRevLett.134.162503},
  url = {https://journals.aps.org/prl/abstract/10.1103/PhysRevLett.134.162503},
  publisher={APS}
}

@article{prc054318 ,
  title={Giant halos in relativistic and nonrelativistic approaches},
  author = {J. Terasaki1 and S. Q. Zhang and S. G. Zhou and J. Meng},
  journal={Phys. Rev. C},
  volume={74},
  pages={054318},
  year={2006},
  doi = {https://doi.org/10.1103/PhysRevC.74.054318},
  url = {https://journals.aps.org/prc/abstract/10.1103/PhysRevC.74.054318},
  publisher={APS}
}

@article{prc047302  ,
  title={Particles in classically forbidden areas, neutron skin and halo, and pure neutron matter
in Ca isotopes},
  author = {Soojae Im and J. Meng},
  journal={Phys. Rev. C},
  volume={61},
  pages={047302 },
  year={2000},
  doi = {https://doi.org/10.1103/PhysRevC.61.047302},
  url = {https://journals.aps.org/prc/abstract/10.1103/PhysRevC.61.047302},
  publisher={APS}
}

@article{prc064319,
  title={Impact of intrinsic electromagnetic structure on the nuclear charge radius
in relativistic density functional theory},
  author = {Hui Hui Xie and Jian Li},
  journal={Phys. Rev. C},
  volume={110},
  pages={064319},
  year={2024},
  doi = {https://doi.org/10.1103/PhysRevC.110.064319},
  url = {https://journals.aps.org/prc/abstract/10.1103/PhysRevC.110.064319},
  publisher={APS}
}

@article{prcL021303,
  title={Charge radii of calcium isotopes within relativistic density functional theory: The ﬁnite size of the nucleon and quadrupole shape-ﬂuctuation effects},
  author = {Hui Hui Xie and Jian Li and Y. L. Yang and P. W. Zhao},
  journal={Phys. Rev. C},
  volume={112},
  pages={L021303},
  year={2025},
  doi = {https://doi.org/10.1103/1kbz-rx3m},
  url = {https://journals.aps.org/prc/abstract/10.1103/1kbz-rx3m},
  publisher={APS}
}

@article{prl092502,
  title={First Observation of Electron Scattering from Online-Produced Radioactive Target},
  author={K. Tsukada and Y. Abe and A. Enokizono and T. Goke and M. Hara and Y. Honda and others},
  journal={Phys. Rev. Lett.},
  volume={131},
  pages={092502},
  year={2023},
  doi = {https://doi.org/10.1103/PhysRevLett.131.092502},
  url = {https://journals.aps.org/prl/abstract/10.1103/PhysRevLett.131.092502},
  publisher={APS}
}

@article{cpc3643,
  title={The neutron halo structure of $^{17}\text{B}$ studied with the relativistic Hartree-Bogoliubov theory},
  author={Juan Xia Ji and Jia Xing Li and Rui Han and Jian Song Wang and Qiang Hu},
  journal={Chinese Phys. C},
  volume={36},
  pages={43},
  year={2012},
  doi = {https://doi.org/10.1088/1674-1137/36/1/007},
  url = {https://iopscience.iop.org/article/10.1088/1674-1137/36/1/007},
  publisher={IOPscience}
}

@article{prl773963,
  title={Relativistic Hartree-Bogoliubov Description of the Neutron Halo in $^{11}\text{Li}$},
  author={J. Meng and P. Ring},
  journal={Phys. Rev. Lett.},
  volume={77},
  pages={3963},
  year={1996},
  doi = {https://doi.org/10.1103/PhysRevLett.77.3963},
  url = {https://journals.aps.org/prl/abstract/10.1103/PhysRevLett.77.3963},
  publisher={APS}
}

@article{prc011301,
  title={Neutron halo in deformed nuclei},
  author = {Shan Gui Zhou and Jie Meng and P. Ring and En Guang Zhao},
  journal={Phys. Rev. C},
  volume={82},
  pages={011301},
  year={2010},
  doi = {https://doi.org/10.1103/PhysRevC.82.011301},
  url = {https://journals.aps.org/prc/abstract/10.1103/PhysRevC.82.011301},
  publisher={APS}
}

@article{prc024312,
  title={Deformed relativistic Hartree-Bogoliubov theory in continuum},
  author = {Lulu Li and Jie Meng and P. Ring and En Guang Zhao and Shan Gui Zhou},
  journal={Phys. Rev. C},
  volume={85},
  pages={024312},
  year={2012},
  doi = {https://doi.org/10.1103/PhysRevC.85.024312},
  url = {https://journals.aps.org/prc/abstract/10.1103/PhysRevC.85.024312},
  publisher={APS}
}

@article{cpl2012294,
  title={Odd Systems in Deformed Relativistic Hartree Bogoliubov Theory in Continuum},
  author = {Lulu Li and Jie Meng and P. Ring and En Guang Zhao and Shan Gui Zhou},
  journal={Chin. Phys. Lett.},
  volume={29},
  pages={042101},
  year={2012},
  doi = {https://doi.org/10.1088/0256-307X/29/4/042101},
  url = {https://cpl.iphy.ac.cn/article/doi/10.1088/0256-307X/29/4/042101},
  publisher={APS}
}

@article{npa600371,
  title={3D solution of Hartree-Fock-Bogoliubov equations for drip-line nuclei},
  author={J. Terasaki and P. H. Heenen and H. Flocard and P. Bonche},
  journal={Nucl. Phys. A},
  volume={600},
  pages={371},
  year={1996},
  doi = {https://doi.org/10.1016/0375-9474(96)00036-X},
  url = {https://www.sciencedirect.com/science/article/pii/037594749600036X?via%3Dihub},
  publisher={ELSEVIER}
}

@article{prc064305,
  title={Nuclear halos and drip lines in symmetry-conserving continuum Hartree-Fock-Bogoliubov theory},
  author = {N. Schunck and J. L. Egido},
  journal={Phys. Rev. C},
  volume={78},
  pages={064305},
  year={2008},
  doi = {https://doi.org/10.1103/PhysRevC.78.064305},
  url = {https://journals.aps.org/prc/abstract/10.1103/PhysRevC.78.064305},
  publisher={APS}
}

@article{npa957416,
  title={Asymptotics of three-body bound state radial wave functions of halo nuclei involving two charged particles},
  author={R. Yarmukhamedov},
  journal={Nucl. Phys. A},
  volume={957},
  pages={416},
  year={2017},
  doi = {https://doi.org/10.1016/j.nuclphysa.2016.10.002},
  url = {https://www.sciencedirect.com/science/article/pii/S0375947416302652},
  publisher={ELSEVIER}
}

@article{PPNP57470,
  title={Relativistic continuum Hartree Bogoliubov theory for ground-state properties of exotic nuclei},
  author={J. Meng and H. Toki and S.G. Zhou and S.Q. Zhang and W.H. Long and L.S. Geng},
  journal={Prog. Part. Nucl. Phys.},
  volume={57},
  pages={470},
  year={2006},
  doi = {https://doi.org/10.1016/j.ppnp.2005.06.001},
  url = {https://www.sciencedirect.com/science/article/pii/S014664100500075X},
  publisher={ELSEVIER}
}

@article{prl082501,
  title={Quasifree Neutron Knockout Reaction Reveals a Small $s$-Orbital Component in the Borromean Nucleus $^{17}\text{B}$},
  author={Z.H. Yang and Y. Kubota and A. Corsi and K. Yoshida and X. X. Sun and others},
  journal={Phys. Rev. Lett.},
  volume={126},
  pages={082501},
  year={2021},
  doi = {https://doi.org/10.1103/PhysRevLett.126.082501},
  url = {https://journals.aps.org/prl/abstract/10.1103/PhysRevLett.126.082501},
  publisher={APS}
}

@article{plb138792,
  title={Nuclear magnetism in the deformed halo nucleus $^{31}\text{Ne}$},
  author={Cong Pan and Kaiyuan Zhang and Shuangquan Zhang},
  journal={Phys. Lett. B},
  volume={855},
  pages={138792},
  year={2024},
  doi = {https://doi.org/10.1016/j.physletb.2024.138792},
  url = {https://www.sciencedirect.com/science/article/pii/S0370269324003502?via%3Dihub},
  publisher={Elsevier}
}

@article{plb138112,
  title={Missed prediction of the neutron halo in $^{37}\text{Mg}$},
  author={K.Y. Zhang and S.Q. Yang and J.L. An and S.S. Zhang},
  journal={Phys. Lett. B},
  volume={844},
  pages={138112},
  year={2023},
  doi = {https://doi.org/10.1016/j.physletb.2023.138112},
  url = {https://www.sciencedirect.com/science/article/pii/S037026932300446X?via%3Dihub},
  publisher={Elsevier}
}

@article{prcL041301,
  title={Possible neutron halo in the triaxial nucleus $^{42}\text{Al}$},
  author = {K. Y. Zhang and S. Q. Zhang and J. Meng},
  journal={Phys. Rev. C},
  volume={108},
  pages={L041301},
  year={2023},
  doi = {https://doi.org/10.1103/PhysRevC.108.L041301},
  url = {https://journals.aps.org/prc/abstract/10.1103/PhysRevC.108.L041301},
  publisher={APS}
}

@article{prc044308,
  title={Triaxial relativistic Hartree-Bogoliubov theory in continuum for exotic nuclei},
  author={K. Y. Zhang and S. Q. Zhang and J. Meng},
  journal={Phys, Rev. C.},
  volume={112},
  pages={044308},
  year={2025},
  doi = {https://doi.org/10.1103/gk2w-cblb},
  url = {https://journals.aps.org/prc/abstract/10.1103/gk2w-cblb},
  publisher={APS}
}

@article{plb139989,
  title={Microscopic description of the proton halo in $^{12}\text{N}$},
  author={K.Y. Zhang and X.X. Lu},
  journal={Phys. Lett. B},
  volume={871},
  pages={139989},
  year={2025},
  doi = {https://doi.org/10.1016/j.physletb.2025.139989},
  url = {https://www.sciencedirect.com/science/article/pii/S0370269325007476},
  publisher={Elsevier}
}

@article{zpa33923,
  title={Relativistic field theory of superfluidity in nuclei},
  author={H. Kucharek and P. Ring},
  journal={Z.Phys.A},
  volume={339},
  pages={23},
  year={1991},
  doi = {https://doi.org/10.1007/BF01282930},
  url = {https://link.springer.com/article/10.1007/BF01282930},
  publisher={Springer}
}

@article{prc024314,
  title={Deformed relativistic Hartree-Bogoliubov theory in continuum with a point-coupling functional: Examples of even-even Nd isotopes},
  author={Kaiyuan Zhang and Myung Ki Cheoun and Yong Beom Choi and Pooi Seong Chong and others},
  journal={Phys, Rev. C.},
  volume={102},
  pages={024314},
  year={2020},
  doi = {https://doi.org/10.1103/PhysRevC.102.024314},
  url = {https://journals.aps.org/prc/abstract/10.1103/PhysRevC.102.024314},
  publisher={APS}
}

@article{prl831112,
  title={Coulomb Dissociation of $^{19}\text{C}$ and its Halo Structure},
  author={T. Nakamura and N. Fukuda and T. Kobayashi and N. Aoi and others},
  journal={Phys. Rev. Lett.},
  volume={83},
  pages={1112},
  year={1999},
  doi = {https://doi.org/10.1103/PhysRevLett.83.1112},
  url = {https://journals.aps.org/prl/abstract/10.1103/PhysRevLett.83.1112},
  publisher={APS}
}

@article{plb39411,
  title={Coulomb excitation of $^{11}\text{Be}$},
  author={T. Nakamura and T. Motobayashi and Y. Ando and A. Mengoni and others },
  journal={Phys. Lett. B},
  volume={394},
  pages={11},
  year={1997},
  doi = {https://doi.org/10.1016/S0370-2693(96)01690-5},
  url = {https://www.sciencedirect.com/science/article/pii/S0370269396016905?via%3Dihub},
  publisher={Elsevier}
}

@article{prl252502,
  title={Observation of Strong Low-Lying E1 Strength in the Two-Neutron Halo Nucleus $^{11}\text{Li}$},
  author={T. Nakamura and N. Aoi and others},
  journal={Phys. Rev. Lett.},
  volume={96},
  pages={252502},
  year={2006},
  doi = { https://doi.org/10.1103/PhysRevLett.96.252502},
  url = {https://journals.aps.org/prl/abstract/10.1103/PhysRevLett.96.252502},
  publisher={APS}
}

@article{prc014316,
  title={Deformed relativistic Hartree-Bogoliubov theory in continuum with a point-coupling functional. II. Examples of odd Nd isotopes},
  author = {Cong Pan and Myung Ki Cheoun and Yong Beom Choi and Jianmin Dong},
  journal={Phys. Rev. C},
  volume={106},
  pages={014316},
  year={2022},
  doi = {https://doi.org/10.1103/PhysRevC.106.014316},
  url = {https://journals.aps.org/prc/abstract/10.1103/PhysRevC.106.014316},
  publisher={APS}
}

@article{prl552676,
  title={Measurements of Interaction Cross Sections and Nuclear Radii in the Light $p$-Shell Region},
  author={I. Tanihata and H. Hamagaki and O. Hashimoto and Y. Shida and others},
  journal={Phys. Rev. Lett.},
  volume={55},
  pages={2627},
  year={1985},
  doi = {https://doi.org/10.1103/PhysRevLett.55.2676},
  url = {https://journals.aps.org/prl/abstract/10.1103/PhysRevLett.55.2676},
  publisher={APS}
}

@article{plb785530,
  title={Shrunk halo and quenched shell gap at $N=16$ in $^{22}$C: Inversion of $sd$ states and deformation effects},
  author={Xiang Xiang Sun and Jie Zhao Shan Gui Zhou},
  journal={Phys. Lett. B},
  volume={785},
  pages={530},
  year={2018},
  doi = {https://doi.org/10.1016/j.physletb.2018.08.071},
  url = {https://www.sciencedirect.com/science/article/pii/S0370269318306907},
  publisher={Elsevier}
}

@article{epja60251,
  title={Toward a unified description of the one-neutron halo nuclei $^{15}$C and 
$^{19}$C from structure to reaction},
  author={Li Yang Wang and Kaiyuan Zhang and Jia Lin An and Shi Sheng Zhang},
  journal={Eur. Phys. J. A},
  volume={60},
  pages={251},
  year={2024},
  doi = {https://doi.org/10.1140/epja/s10050-024-01464-7},
  url = {https://link.springer.com/article/10.1140/epja/s10050-024-01464-7},
  publisher={Springer}
}

@article{adndt101488,
  title={Nuclear mass table in deformed relativistic Hartree–Bogoliubov theory in continuum, I: Even–even nuclei},
  author={Kaiyuan Zhang and Myung Ki Cheoun and Yong Beom Choi and Pooi Seong Chong and others},
  journal={At. Data Nucl. Data Tables},
  volume={144},
  pages={101488},
  year={2022},
  doi = {https://doi.org/10.1016/j.adt.2022.101488},
  url = {https://www.sciencedirect.com/science/article/pii/S0092640X22000018?via%3Dihub},
  publisher={ELSEVIER}
}

@article{adndt101661,
  title={Nuclear mass table in deformed relativistic Hartree–Bogoliubov theory in continuum, II: Even-$Z$ nuclei},
  author={Peng Guo and Xiaojie Cao and Kangmin Chen and Zhihui Chen and others},
  journal={At. Data Nucl. Data Tables},
  volume={158},
  pages={101661},
  year={2024},
  doi = {https://doi.org/10.1016/j.adt.2024.101661},
  url = {https://www.sciencedirect.com/science/article/pii/S0092640X24000263?via%3Dihub},
  publisher={ELSEVIER}
}

@article{prcL021301,
  title={Predictive power for superheavy nuclear mass and possible stability beyond the neutron drip line in deformed relativistic Hartree-Bogoliubov theory in continuum},
  author = {Kaiyuan Zhang and Xiaotao He and Jie Meng and Cong Pan and others},
  journal={Phys. Rev. C},
  volume={104},
  pages={L021301},
  year={2021},
  doi = {https://doi.org/10.1103/PhysRevC.104.L021301},
  url = {https://journals.aps.org/prc/abstract/10.1103/PhysRevC.104.L021301},
  publisher={APS}
}

@article{prc014301,
  title={Odd-even differences in the stability “peninsula” in the $106 \leq \text{Z} \leq 112$ region with the deformed relativistic Hartree-Bogoliubov theory in continuum},
  author = {Xiao Tao He and Jia Wei Wu and Kai Yuan Zhang and Cai Wan Shen},
  journal={Phys. Rev. C},
  volume={110},
  pages={014301},
  year={2024},
  doi = {https://doi.org/10.1103/PhysRevC.110.014301},
  url = {https://journals.aps.org/prc/abstract/10.1103/PhysRevC.110.014301},
  publisher={APS}
}

@article{AAPPSbull3513,
  title={Selected advances in nuclear mass predictions based on covariant density functional theory with continuum effects},
  author={K. Y. Zhang and C. Pan and X. H. Wu and X. Y. Qu and X. X. Lu and G. A. Sun},
  journal={AAPPS Bull.},
  volume={35},
  pages={13},
  year={2025},
  doi = {https://doi.org/10.1007/s43673-025-00153-x},
  url = {https://link.springer.com/article/10.1007/s43673-025-00153-x},
  publisher={Springer}
}

@article{nst36231,
  title={Benchmarking nuclear energy density functionals with new mass data},
  author={Xiao Ying Qu and Kang Min Chen and Cong Pan and Yang Yang Yu and Kai Yuan Zhang},
  journal={Nucl. Sci. Tech.},
  volume={36},
  pages={231},
  year={2025},
  doi = {https://doi.org/10.1007/s41365-025-01821-1},
  url = {https://link.springer.com/article/10.1007/s41365-025-01821-1},
  publisher={Springer}
}

@article{prc0143252,
  title={Effective field theory for proton halo nuclei},
  author = {Emil Ryberg and Christian Forssén and H. W. Hammer and Lucas Platter},
  journal={Phys. Rev. C},
  volume={89},
  pages={014325},
  year={2014},
  doi = {https://doi.org/10.1103/PhysRevC.89.014325},
  url = {https://journals.aps.org/prc/abstract/10.1103/PhysRevC.89.014325},
  publisher={APS}
}

@article{prc044004,
  title={$^{6}\text{He}$ nucleus in halo effective field theory},
  author = {C. Ji and Ch. Elster and D. R. Phillips},
  journal={Phys. Rev. C},
  volume={90},
  pages={044004},
  year={2014},
  doi = {https://doi.org/10.1103/PhysRevC.90.044004},
  url = {https://journals.aps.org/prc/abstract/10.1103/PhysRevC.90.044004},
  publisher={APS}
}

@article{JPG103002,
  title={Effective field theory description of halo nuclei},
  author={H W Hammer and C Ji and D R Phillips},
  journal={J. Phys. G},
  volume={44},
  pages={103002},
  year={2017},
  doi = {10.1088/1361-6471/aa83db},
  url = {https://iopscience.iop.org/article/10.1088/1361-6471/aa83db},
  publisher={IOPscience}
}

@article{prc034305,
  title={Halo nuclei $^{6}\text{He}$ and $^{8}\text{He}$ with the Coulomb-Sturmian basis},
  author = {M. A. Caprio and P. Maris and J. P. Vary},
  journal={Phys. Rev. C},
  volume={90},
  pages={034305},
  year={2014},
  doi = {https://doi.org/10.1103/PhysRevC.90.034305},
  url = {https://journals.aps.org/prc/abstract/10.1103/PhysRevC.90.034305},
  publisher={APS}
}

@article{prc061302,
  title={Ab initio Gamow in-medium similarity renormalization group with resonance and continuum},
  author = {B. S. Hu and Q. Wu and Z. H. Sun and F. R. Xu},
  journal={Phys. Rev. C},
  volume={99},
  pages={061302},
  year={2019},
  doi = {https://doi.org/10.1103/PhysRevC.99.061302},
  url = {https://journals.aps.org/prc/abstract/10.1103/PhysRevC.99.061302},
  publisher={APS}
}

@article{prl2425012,
  title={an Ab Initio Theory Explain the Phenomenon of Parity Inversion in $^{11}\text{Be}$?},
  author={Angelo Calci1 and Petr Navrátil and Robert Roth and Jérémy Dohet-Eraly and Sofia Quaglioni and Guillaume Hupin},
  journal={Phys. Rev. Lett.},
  volume={117},
  pages={242501},
  year={2016},
  doi = {https://doi.org/10.1103/PhysRevLett.117.242501},
  url = {https://journals.aps.org/prl/abstract/10.1103/PhysRevLett.117.242501},
  publisher={APS}
}

@article{prcL061304,
  title={Unveiling potential neutron halos in intermediate-mass nuclei: An ab initio study},
  author = {H. H. Li and J. G. Li and M. R. Xie and W. Zuo},
  journal={Phys. Rev. C},
  volume={109},
  pages={L061304},
  year={2024},
  doi = {https://doi.org/10.1103/PhysRevC.109.L061304},
  url = {https://journals.aps.org/prc/abstract/10.1103/PhysRevC.109.L061304},
  publisher={APS}
}

@article{adndt9969,
  title={Table of experimental nuclear ground state charge radii: An update},
  author={I. Angeli and K.P. Marinova},
  journal={At. Data Nucl. Data Tables},
  volume={99},
  pages={69},
  year={2013},
  doi = {https://doi.org/10.1016/j.adt.2011.12.006},
  url = {https://www.sciencedirect.com/science/article/pii/S0092640X12000265},
  publisher={Elsevier}
}

@article{prc0443082,
  title={Nuclear ground state observables and QCD scaling in a refined relativistic point coupling model},
  author = {T. Bürvenich and D. G. Madland and J. A. Maruhn and P.-G. Reinhard},
  journal={Phys. Rev. C},
  volume={65},
  pages={044308},
  year={2002},
  doi = {https://doi.org/10.1103/PhysRevC.65.044308},
  url = {https://journals.aps.org/prc/abstract/10.1103/PhysRevC.65.044308},
  publisher={APS}
}

@article{plb137946,
  title={The optimized point-coupling interaction for the relativistic energy density functional of Hartree-Bogoliubov approach quantifying the nuclear bulk properties},
  author={Zi Xin Liu and Yi Hua Lam and Ning Lu and and Peter Ring},
  journal={Phys. Lett. B},
  volume={842},
  pages={137946},
  year={2023},
  doi = {https://doi.org/10.1016/j.physletb.2023.137946},
  url = {https://www.sciencedirect.com/science/article/pii/S0370269323002800?via%3Dihub},
  publisher={Elsevier}
}

@article{prc044304 ,
  title={Neutron scattering off one-neutron halo nuclei in halo effective field theory},
  author = {Xu Zhang and Hai Long Fu and Feng-Kun Guo and Hans Werner Hammer},
  journal={Phys. Rev. C},
  volume={108},
  pages={044304},
  year={2023},
  doi = {https://doi.org/10.1103/PhysRevC.108.044304},
  url = {https://journals.aps.org/prc/abstract/10.1103/PhysRevC.108.044304},
  publisher={APS}
}

@article{prl2225012,
  title={How Many-Body Correlations and $\alpha$ Clustering Shape $^{6}\text{He}$},
  author={Carolina Romero-Redondo and Sofia Quaglioni and Petr Navrátil and Guillaume Hupin},
  journal={Phys. Rev. Lett.},
  volume={117},
  pages={222501},
  year={2016},
  doi = {https://doi.org/10.1103/PhysRevLett.117.222501},
  url = {https://journals.aps.org/prl/abstract/10.1103/PhysRevLett.117.222501},
  publisher={APS}
}

@article{prc034323 ,
  title={Spherical relativistic Hartree theory in a Woods-Saxon basis},
  author = {Shan-Gui Zhou and Jie Meng and P. Ring},
  journal={Phys. Rev. C},
  volume={68},
  pages={034323},
  year={2003},
  doi = {https://doi.org/10.1103/PhysRevC.68.034323},
  url = {https://journals.aps.org/prc/abstract/10.1103/PhysRevC.68.034323},
  publisher={APS}
}

@article{prc024302 ,
  title={Optimized Dirac Woods-Saxon basis for covariant density functional theory},
  author = {K. Y. Zhang and C. Pan and S. Q. Zhang},
  journal={Phys. Rev. C},
  volume={106},
  pages={024302},
  year={2022},
  doi = {https://doi.org/10.1103/PhysRevC.106.024302},
  url = {https://journals.aps.org/prc/abstract/10.1103/PhysRevC.106.024302},
  publisher={APS}
}

@article{prc034319,
  title={New effective interactions in relativistic mean field theory with nonlinear terms and density-dependent meson-nucleon coupling},
  author = {Wenhui Long and Jie Meng and Nguyen Van Giai and Shan Gui Zhou},
  journal={Phys. Rev. C},
  volume={69},
  pages={034319},
  year={2004},
  doi = {https://doi.org/10.1103/PhysRevC.69.034319},
  url = {https://journals.aps.org/prc/abstract/10.1103/PhysRevC.69.034319},
  publisher={APS}
}

@article{nst33153,
  title={Prediction of nuclear charge density distribution with feedback neural network},
  author={Tian Shuai Shang and Jian Li and Zhong Ming Niu},
  journal={Nucl. Sci. Tech.},
  volume={33},
  pages={153},
  year={2022},
  doi = {https://doi.org/10.1007/s41365-022-01140-9},
  url = {https://link.springer.com/article/10.1007/s41365-022-01140-9},
  publisher={Springer}
}

@article{prc014308,
  title={Global prediction of nuclear charge density distributions using a deep neural network},
  author = {Tian Shuai Shang and Hui Hui Xie and Jian Li and Haozhao Liang},
  journal={Phys. Rev. C},
  volume={110},
  pages={014308},
  year={2024},
  doi = {https://doi.org/10.1103/PhysRevC.110.014308},
  url = {https://journals.aps.org/prc/abstract/10.1103/PhysRevC.110.014308},
  publisher={APS}
}

@article{nst3793,
  title={Predictions of charge density distributions for nuclei with Z $\geq$ 8},
  author={Yun Dong Wang and Tian Shuai Shang and Hui Hui Xie and Peng Xiang Du and Jian Li and Hao Zhao Liang},
  journal={Nucl. Sci. Tech.},
  volume={37},
  pages={93},
  year={2026},
  doi = {https://doi.org/10.1007/s41365-026-01905-6},
  url = {https://link.springer.com/article/10.1007/s41365-026-01905-6},
  publisher={Springer}
}

\end{document}